\documentclass[preprint, aps, pre]{revtex4-1}
\pdfoutput=1
\usepackage[utf8]{inputenc}
\usepackage{amsmath}
\usepackage{amsfonts}
\usepackage{graphicx}
\usepackage{amssymb}
\usepackage{hyperref}
\usepackage[normalem]{ulem}
\usepackage{subcaption}
\usepackage[dvipsnames]{xcolor}

\begin{document}
\title{Effects of orientational order on modulated cylindrical interfaces}
\author{Jason Klebes$^{*\dagger}$}
\author{Paul Clegg$^\dagger$}
\author{R M L \surname{Evans$^*$}}
\affiliation{$^*$School of Mathematics, University of Leeds, Leeds LS2 9JT, United Kingdom}
\affiliation{$^\dagger$School of Physics and Astronomy, University of Edinburgh, James Clerk Maxwell Building, Peter Guthrie Tate Road, Edinburgh, EH9 3FD, United Kingdom}
\begin{abstract}
Cylindrical interfaces occur in sheared or deformed emulsions and as biological or technological lipid monolayer or bilayer tubules.  Like the corresponding spherical droplets and vesicles, these cylinder-like surfaces may host orientaional order with $n$-fold rotational symmetry, for example in the positions of lipid molecules or of spherical nanoparticles.  We examine how that order interacts with and induces shape modulations of cylindrical interfaces.  While on spherical droplets $2n$ topological defects necessarily exist and can induce icosahedral droplet shapes, the cylindrical topology is compatible with a defect-free patterning.  Nevertheless, once a modulation is introduced by a mechanism such as spontaneous curvature, nontrivial patterns of order, including ones with excess defects, emerge and have nonlinear effects on the shape of the tube.  Examining the equilibrium energetics of the system analytically and with a lattice-based Markov chain Monte Carlo simulation, we predict low-temperature morphologies of modulated cylindrical interfaces hosting orientational order.  A shape modulation induces a banded pattern of alternatingly isotropic and ordered interfacial material.  Furthermore cylindrical systems can be divided into Type I, without defects, and Type II, which go through a spectrum of defect states with up to $4n$ excess defects. The character of the curvature-induced shape transition from unmodulated to modulated cylinders is continuous or discontinuous accordingly.
\end{abstract}
\maketitle
\section{Introduction}
The ubiquitous emulsion is a metastable material where droplets of one fluid are suspended in another. All emulsions in practical use require surfactants in the form of amphiphilic molecules or, in the case of solid-stabilized or Pickering emulsions, in the form of nano- or miocroparticles.   In both cases the surfactants form an interfacial layer, which extends the emulsion lifetime:  molecular surfactants greatly reduce surface tension, while in solid-stabilized emulsions superlative stability is achieved by particles' strong adsorption to interfaces \cite{binks2002}.
However, the interfacial layer is itself a complex material of interacting molecules or particles, loosely confined to a two-dimensional curved interface.  Beyond the desired stabilizing effect, the interfacial layer can impart spatially varying mechanical properties onto the interface and influence the morphology of emulsion droplets.  

In common experience, a droplet minimizing its surface area adopts a perfectly spherical shape.
While emulsion droplets usually conform to this expectation, at low temperatures, the surface tension can vanish and effects from a hexatically ordered interfacial layer dominate, inducing polyhedral droplets \cite{marin2020, guttman2016}.  Factors such as a negative surface tension and gravity \cite{garcia-aguilar2021} induce further exotic droplet morphologies such as flattened polygons, rods, and protrusions. Solid-stabilized emulsion, whose stabilization mechanism differs, can form similarly off-spherical facetted shapes at room temperature \cite{abkarian2007}.  Despite differences in energy scales, length scales, and driving mechanisms, the icosahedral facetted morphology in several experimental systems is induced by the interaction of hexatic order with a spherical surface topology.  The positions of particles or molecules on a two-dimensional surface will, at low temperature, arrange in a way locally resembling a hexagonal close packing.  The phase, with quasi-long-range correlation in the orientation of the pattern, is known as hexatic.  Just as a sphere cannot be covered by a vector field without two point-defects, where the vector field diverges, it cannot be smoothly covered by hexatic orientational order.  The necessary total number of defects is commonly realized in the form of twelve defect sites, where a particle has only five neighbors, in an icosohedral arrangement on the sphere.  The prescence of twelve topologically induced defects can be seen directly when low-temperature droplets adopt a facetted icosahedral morphology \cite{guttman2016}.

In contrast, order on a cylindrical surface has no topologically mandatory defects.  While uniform hexatic order on a cylindrical surface has a complex effect on instability and dynamics, as studied by Lenz and Nelson \cite{lenz2003}, no phenomenon comparable to faceting is predicted in the linear analysis.  Examples of cylinder-like systems with an interfacial layer or membrane are biological lipid nanotubes and their nanotechnological counterpart\cite{eevans1996}.  Long tails are also seen to grow from the above cooled emulsion droplets \cite{guttman2016}.  An experimental system of a larger lipid bilayer tube with a dynamic instability was introduced by Bar-Ziv and Moses \cite{bar-ziv1994}.  Cylinder-like geometries also occur transiently in most industrial emulsions during the formation process \cite{stone1994}.  As a result of arrested coalescence, modulated cylindrical structures can form and persist in emulsions.  Through specialized mixing techniques, solid-stabilized emulsions with long-lived elongated morphologies can be manufactured \cite{li2019}.

Observing that in a variety of cylindrical systems, a modulated or pearled morphology is induced by factors such as spontaneous curvature or external forces, we examine whether excess defects will appear conditionally on modulated cylinders.  As on toroidal droplets \cite{evans1995, bowick2009}, the appearance of `excess' defects in charge-neutral pairs on cylindrical systems is possible, but not topologically inevitable.  A modulated cylindrical shape is seen in a variety of cylinder-like systems, possibly including in the coiled tails of facetted emulsions droplets themselves. Spontaneous curvature has been proposed as a mechanism behind the instability of lipid bilayer tubules as well as cylindrical surfactant micelles \cite{granek1996,chaieb1998}. Isotropic spontaneous curvature can exist due to an asymmetry of layers in a lipid bilayer or due to the geometry of lipid molecules or additives in a lipid monolayer.
We here use an off-neutral spontaneous curvature as a representative initial driver of a modulated morphology. 

First, we establish the energetics of cylinder-like systems dominated by spontaneous curvature only.  To see how $n$-atic order (the generalization of hexatic order to order with $n$-fold rotational symmetry) modifies the behavior, we then use a field-theoretic approach to in-plane orientational order to obtain configurations and energetics of $n$-atic order on modulated cylinders.  This approach allows us to reveal a polymorphic spectrum of morphologies with different numbers of defects.

\section{Geometry and orientational order}
Our model surface is a cylinder with a sinusoidal modulation in radius.
In three-dimensional space spanned by a cylindrical polar coordinate basis  ($\bf{e}_{\rho}$, $\bf{e}_{\theta}$, $\bf{e}_{z}$), the surface of revolution is described by the parametric equation

\begin{equation}
\begin{aligned}
\label{eq:surface}
\bf{x}(\theta, z)&= r(a) (1+a \sin (kz)) \bf{e}_{\rho}+ z \bf{e}_{z} \\
\end{aligned}
\end{equation} 

The dimensionless parameter $a \in (-1,1)$, the shape amplitude, gives the amplitude of the sinusoidal modulation.  The surfaces are subject to a global volume-conserving constraint, $V(a) = V_0$, due to the incompressible inner fluid.  Consequently the mean radius $r(a)$ in equation \ref{eq:surface} must depend on shape amplitude as 
\begin{equation}
\label{radiusrescaled}
r(a) = \dfrac{r_0}{\sqrt{1+a^2/2}}.
\end{equation}

At $a=0$ the surface is a flat cylinder of radius $r_0$.  
We define the length unit by setting $r_0=1$.  The sinusoidal modulation is additionally characterized by its wavenumber $k$ and wavelength $\lambda=2\pi/k$. 

\begin{figure}
\centering
\includegraphics[width=.45\textwidth]{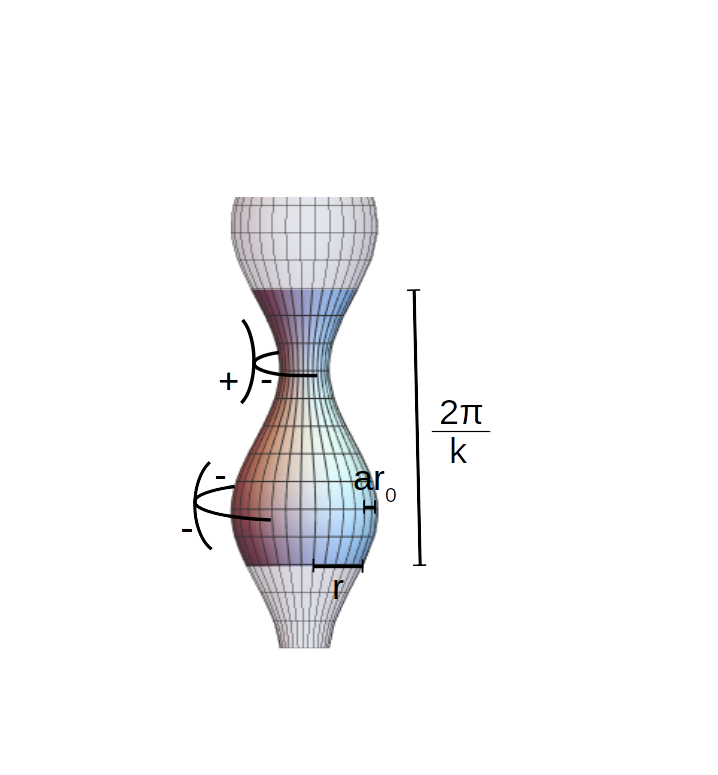}
\caption{The model surface, a sinusoid periodic surface of revolution.  Positive and negative principal curvatures occur.  The sinusoidal modulation has amplitude $ar_0$ and wavelength $\lambda= 2\pi/k$.   The mean radius $r(a)$ is chosen to conserve volume relative to a cylinder of radius $r_0$. }
\end{figure}

We assume the system is closed and periodic, neglecting the topological effects of end-caps or attachment to larger spheroids that may be present in a real emulsion tubule. It thus has topological genus 1, in common with the torus.  While a topological sphere has $2n$ topologically mandated defects, the periodic model system has none. In comparable elongated closed vesicles, the $2n$ defects are commonly localized at the end-caps \cite{mesarec2017}.  Our periodic system, omitting both end-caps and their defects, facilitates the study of excess defect pairs in isolation.  Alternatively our periodic model can represent a bridge between larger fluid reservoirs in a material with a more complex topology, whose Gaussian curvature is similarly omitted along with the associated defects.

The external Hamiltonian $\mathcal{H}_E $ describes the energy of the membrane or interface, that is the part of the system energy not due to internal degrees of freedom of the interfacial layer.  The Helfrich\cite{helfrich1973} Hamiltonian is a widely used estimate of the energetics of cell membranes (lipid bilayers) with surface tension and bending rigidity:
\begin{equation}
\label{eq:hh}
\mathcal{H}_E = \gamma_0 \int_S dS + \frac{\kappa}{2}\int_S (2H-H_0)^2 dS+ \frac{\bar{\kappa}}{2}\int_S K dS
\end{equation}
with $\gamma_0$ a general microscopic surface tension, $H$ mean curvature, $H_0$ the material's spontaneous total curvature, $K$ Gaussian curvature, and $\kappa$  and $\bar{\kappa}$ two bending rigidities.  

All integrals are over one period of the surface, $\int_S dS = \int_0^{\lambda} \int_0^{2\pi} d\theta dz \sqrt{g}$.  The square root of the metric determinant, $\sqrt{g}=\sqrt{g_{zz} g_{\theta \theta}}$, can be seen as the measure of the integral, describing the relative size of an infinitesimal area element. The metric tensor, in the basis of the cylindrical coordinate system, can be calculated as $g_{ij}= \partial_i \bf{x} \cdot \partial_j \bf{x}$:
\begin{equation}
\label{eq:g}
\begin{aligned}
g_{\theta \theta} &=  r(a)^2 (1+a \sin(kz))^2\\
g_{zz} &= 1+a^2k^2\cos^2(kz)\\
g_{\theta z} &= g_{z \theta}=0.\\
\end{aligned}
\end{equation}

We will additionally use the shape tensor $K^i_j$. When diagonal, it gives the two  principal curvatures at each point.  It can be derived as $K_{ij}=\hat{n} \cdot \partial_i \partial_j \bf{x}$, $K^i_j = K_{kj}g^{ki}$, where $n$ is the unit normal to the surface and the index-raising $g^{ij}$ is the inverse of the metric tensor.  Here the principal curvatures are
\begin{equation}
\begin{aligned}
\label{eq:curvaturetensor}
&K_{\theta}^{\theta}=\dfrac{-1}{ \sqrt{g}}\\
&K_{z}^{z}=\dfrac{-r(a) a k^2\sin(kz)}{ (g_{zz})^{3/2}}
\end{aligned}
\end{equation}
Since $K^i_j$ is diagonal in our basis, mean curvature $H$ and Gaussian curvature $K$ are given by its trace and determinant as $2H= K_\theta^\theta +K_z^z$ and $K=K_\theta^\theta K_z^z$ respectively.

For further discussion of geometric quantities and the Helfrich Hamiltonian, see review \cite{kamien2002}.

We here examine a surface layer with $n$-atic orientational order.  The $n$-atic order parameter represents correlations in, for example, the 1-atic vector direction of the tilt of lipid molecules, 2-atic orientation of a nematic liquid crystal layer, or hexatic order in the orientation of hexagonal arrangements of the positions of particles.  Order is described by order parameter field $\Psi(\bf{x})$, which holds information on $n$-atic orientational order of molecules or particles which make up the interfacial layer.  

From the local structure of the interfacial layer, the complex-valued order parameter field is defined as

\begin{equation}
\Psi(\bf{x})= \langle  e^{i n \phi(\bf{x})} \rangle .
\label{eq:psidef}
\end{equation}

The angle $\phi(\bf{x})$ indicates orientation of a local molecule or particle.  This may be the tilt of a molecule ($n=1$), the orientation of a rod-like molecule or particle (nematic, $n=2$), or the direction of an imagined bond between the positions of a particle and its neighbor in a hexatic arrangement ($n=6$).  The factor $n$ is the order of the discrete rotational symmetry of the material.  The brackets $\langle \rangle$ denote a local spatial averaging.  The magnitude as well as the phase of $\Psi(\bf{x})$ may vary, with $|\Psi|=0$ corresponding to the isotropic state.

The angle $\phi$ in equation \ref{eq:psidef} is defined with respect to one arbitrarily chosen axis of an intrinsic coordinate system spanned by the unit tangent vectors $\bf{t}$ to the surface.  They are, in the basis $\bf{e}_i$, 
\begin{equation}
\label{eq:tangentvectors}
\bf{t}_i= \frac{\partial_i \bf{x}}{|\partial_i \bf{x}|}.
\end{equation}

The order parameter field is subject to a Ginzburg-Landau potential, the simplest analytic expression reproducing the desired phase behavior and coupling to surface curvature
\begin{equation}
\label{eq:hi}
\mathcal{H}_I= \int_S dS \left( \alpha |\Psi|^2 +  c |D_i\Psi|^2 + \frac{u}{2}|\Psi|^4\right) .
\end{equation}
Coefficient $\alpha$ is a temperature-dependent material parameter which is negative below the critical temperature; and $c$ and $u$ are positive parameters.  The Landau-Ginzburg model for $n$-atic order of membranes on curved interfaces has been introduced for spheres by Park et al. \cite{park1992} and studied on spheres and tori by Evans \cite{evans1995, evans1996}.  For a summary of further developments in the study of Landau-Ginzburg $n$-atic order and of other representations of order on a variety of surface shapes, see review by Bowick and Giomi \cite{bowick2009}.

The field magnitude $|\Psi|$ is sometimes taken to be constant except for point defects, so that only the gradient energy varies, for example by Lenz and Nelson \cite{lenz2003}, treating the onset of instabilities of spheres and cylinders with hexatic order, or in a treatment by Kumaralageshan et al. \cite{kumaralageshan2017} which showcases the elegant differential geometric solutions enabled by the assumption.   We here retain varying magnitude.  The more general theory allows us to represent a high-temperature regime where $\alpha$ is close to the critical value \cite{foltin2000}, at the expense of analytical tractability.

On flat surfaces and in the absence of thermal fluctuations, well-known solutions minimizing the Landau theory, i.e. Equation \ref{eq:hi} without the gradient term, are a spatially constant field $\Psi(\bf{x})=\Psi_0$ with magnitude
\begin{equation}
|\Psi_0| =
    \begin{cases}
            \sqrt{\frac{-\alpha}{u}}, &          \alpha \leq 0,\\
            0, &         \alpha>0.
    \end{cases}
\end{equation}
and arbitrary phase.
In this case the material has an energy density 
\begin{equation}
f_0 =
    \begin{cases}
            -\frac{\alpha^2}{2u}, &          \alpha \leq 0,\\
            0, &         \alpha>0;
    \end{cases}
\end{equation}
the negative free energy density of the ordered interface (relative to surfactants in the bulk) at low temperatures can act as an effective negative surface tension \cite{guttman2016}.
The material has a persistence length \cite{chaikin1995} of
\begin{equation}
\xi =
    \begin{cases}
            \sqrt{\frac{c}{2 |\alpha|}}, &          \alpha \leq 0,\\
            \sqrt{\frac{c}{\alpha}}, &         \alpha>0.
    \end{cases}
    \end{equation}

On curved surfaces, the field is coupled to surface geometry via the covariant derivative operator 

\begin{equation}
D_i = \partial_i - i n A_i,
\end{equation}
where $A_i$ is the spin connection, a quantity related to surface shape at each point.  The spin connection corrects for deviations in parallel transport on curved surfaces, allowing comparison of the vector field at two distant points.  The formulation of the covariant derivative operator,  first introduced in this form for the study of $n$-atic material on curved surfaces by Park et al. \cite{park1992}, multiplies the spin connection by a factor of $i n$ to account for the mapping of $1/n$ of a full turn in orientation to a full rotation of complex phase.  
One way to derive the spin connection is $A_i =\bf{t}_\theta \cdot \partial_i \bf{t}_z$.  Starting from Equation \ref{eq:surface} and retrieving tangent vectors via Equation \ref{eq:tangentvectors}, the spin connection for the given surface is
\begin{equation}
\label{eq:connection}
\begin{aligned}
A_z&=0\\
A_\theta &= \frac{r(a) a k \cos(kz)}{\sqrt{g_{zz}}}.
\end{aligned}
\end{equation}

\section{Spontaneous curvature}
\label{chap:external}
For cylindrical systems dominated by surface tension, $\mathcal{H}_E=\gamma \int_S dS$ only, the well-known Plateau-Rayleigh limit of stability is (in units of $1/r_0$) critical wavenumber $k_c=1$.  The limit of stability can be derived by examining the linearized energy difference associated with a small perturbation.  Sinusoidal perturbations with smaller wavenumber (longer wavelength) than the limiting value decrease the energy of the system and therefore grow.  For our system, the calculation is repeated using Equation \ref{eq:hh} only (in the absence of orientational order described by Equation \ref{eq:hi}).

Our surface is a periodic tube; we will take this model literally and assume it is a closed surface of topological genus $g=1$.
On a closed surface of constant topological genus, the total surface integral of Gaussian curvature is a constant $2\pi\chi$, determined completely by the surface's Euler characteristic $\chi = 2-2g$ (Gauss-Bonnet theorem).  
Thus, the third term in Equation \ref{eq:hh}, relating to total Gaussian curvature, is a constant and will be dropped.

Expanding the second term in Equation \ref{eq:hh}, a cross-term $-2\kappa K_\theta^\theta K_z^z$ is also proportional to the Gaussian curvature, thus its integral is also a constant. The external Hamiltonian is reduced to

\begin{equation}
\begin{aligned}
\label{eq:Hcurv}
\mathcal{H}_E&=\mathcal{H}_E^{surf}+\mathcal{H}_E^{curv}\\
\mathcal{H}_E^{surf}&=\gamma_0 \int_S dS\\
\mathcal{H}_E^{curv}&=\frac{\kappa}{2} \int_S dS ((K_\theta^\theta)^2+(K_z^z)^2 -2K^\theta_\theta H_0 - 2K_z^zH_0 +H_0^2)\\
\end{aligned}
\end{equation}

Inserting expressions from Equations \ref{eq:curvaturetensor} and \ref{eq:g}, we expand all analytic functions in Equation \ref{eq:Hcurv} as series in small $a$, and integrate over one period.  Examining the next-to-leading order term, we retrieve the energy difference on a small sinusoidal perturbation of amplitude $a$ and wavenumber $k$:  

\begin{equation}
\label{eq:diff}
    \dfrac{\Delta \mathcal{H}_E}{A_0 a^2}= \frac{\gamma}{4} \left( k^2 - 1 \right)+  \frac{1}{8} \left(2 k^4+(4 H_0 -1) k^2 +3  \right) ,
\end{equation}
where $A_0= 4 \pi^2 / k$ is the surface area of a section of length $\lambda=2\pi/k $ of the unperturbed, cylindrical surface. The last term in Equation \ref{eq:Hcurv} has been absorbed into the surface tension $\gamma= \gamma_0 + H_0^2/2$ and the equation has been nondimensionalized by choosing units where $\kappa=1$.  

Roots of Equation \ref{eq:diff} are the critical wavenumber

\begin{equation}
\begin{aligned}
\label{eq:kc}
k_c(H_0) &= \frac{1}{2} \Big( 1-2 \gamma- 4 H_0 
\\& \pm \sqrt{8 (2 \gamma -3) + (-1+4 H_0 + 2 \gamma )^2}\Big)^{1/2}.
\end{aligned}
\end{equation}

Equations \ref{eq:diff} and \ref{eq:kc} are known in various forms and special cases in the literature on the pearling instability of lipid bilayer membrane tubules \cite{boedec2014}.

The surface tension is composed of the constant $H_0^2/2$ from equation \ref{eq:Hcurv}, a local energy density $f_0$ of the possibly ordered interfacial material as described by Equation \ref{eq:hi}, and any other effects that may be present, represented by $\gamma_0$.  For simplicity of calculations we neglect additional effects, such that $\gamma_0=0$ and $\gamma:=f_0 +  H_0^2/2$.  In the low-temperature isotropic phase $f_0= 0$ and we have only $\gamma:=  H_0^2/2$:  the effective surface tension is dominated by a constant energy density from spontaneous curvature.

\begin{figure}
\centering
\includegraphics[width=.45\textwidth]{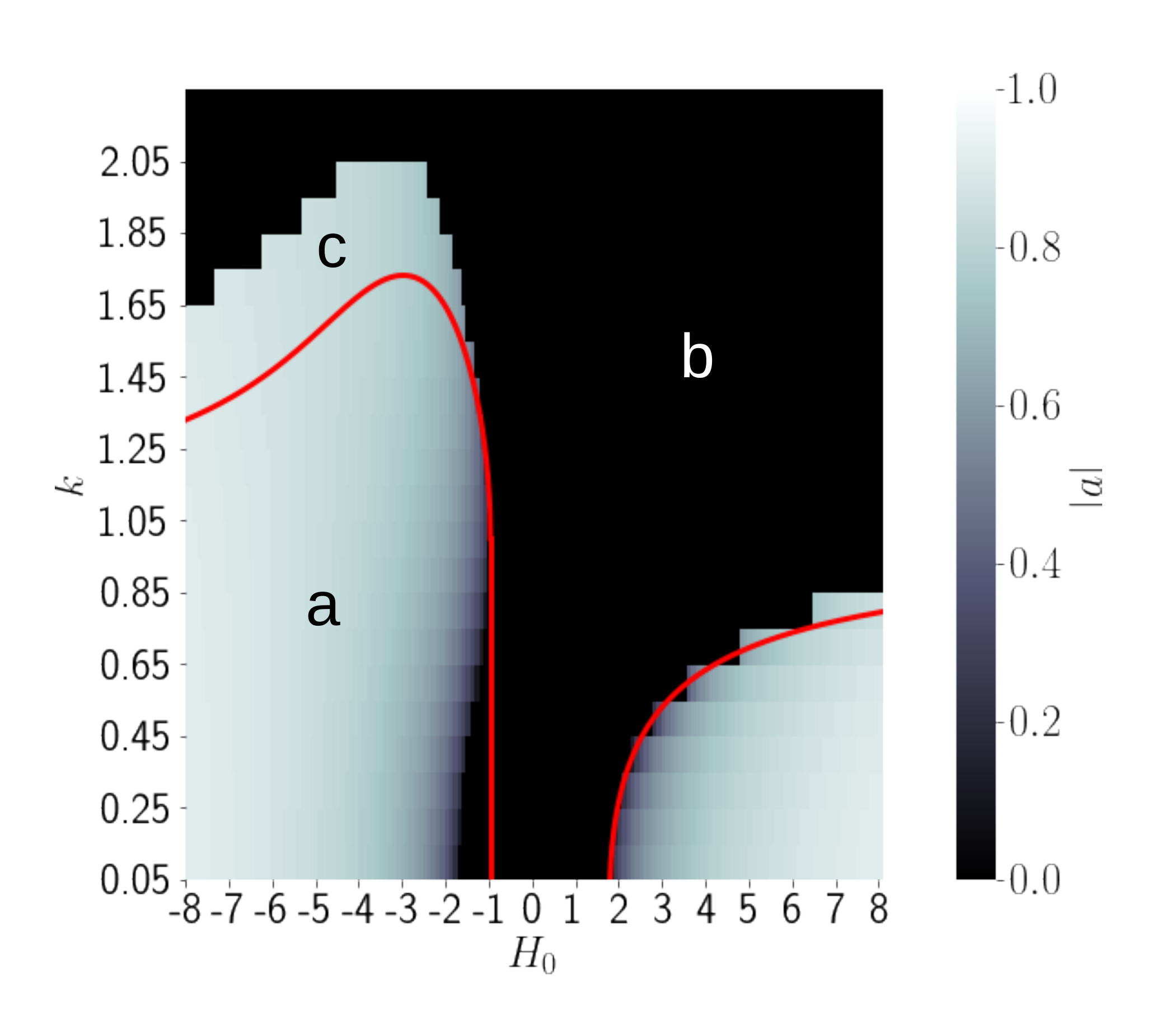}
\caption{ Stability as a function of wavenumber and spontaneous curvature in the absence of orientational order.  The red line $k_c(H_0)$ is the critical wavenumber below which cylinders are linearly unstable according to the perturbative calculation, while background shading indicates shape amplitude $a$ that is the global minimum of $\mathcal{H}_E(a)$ at the given $k$ and $H_0$. (a) Light background shading ($a>0$) and wavenumbers below the critical value $k_c(H_0)$ indicate instability according to both linearly and numerical analysis.  (b) Black background shading ($a=0$) and wavenumers larger than $k_c$ indicate that the unperturbed cylinder is stable according to both indicators.  (c) Where the two indicators of stability disagree, the cylinder is metastable.}
\label{fig:numerical_H0_g0}
\end{figure}

The critical wavenumber as a function of spontaneous curvature in the abscence of orientational order is shown as the line in Figure \ref{fig:numerical_H0_g0}. 
The system is absolutely stable against shape perturbation of all wavelengths at spontaneous curvatures from $H_0=-1$ to $\sqrt{3}$.
While there are two real solutions $k_c(H_0)$ to Equation \ref{eq:diff} for intrinsic curvatures between $H_0=-\sqrt{3}$ to $H_0=-1$, all wavenumbers below the upper curve should be counted as unstable because these long-wavelength systems are unstable against smaller-wavelength shape modulations.

At extreme spontaneous curvature of either sign, the effective surface tension $\gamma =  \kappa H_0^2/2$ dominates and the system approaches the original Plateau-Rayleigh stability criterion, $k_c = 1$. 

Orientation of the surface is defined so that the original cylinder has the negative total curvature $2H = -1/r_0= -1$. Surprisingly, a spontaneous curvature with the same sign and slightly larger magnitude has a more prominent destabilizing effect than a positive spontaneous curvature.  
The maximum critical wavenumber $k_c=\sqrt{3}$ occurs at $H_0 = -3$.  The instability can be explained by considering the axial curvature of a sinusoidal perturbation.  The larger parts of the channel have both principal curvatures negative, approaching sphere-like.  The positive axial principal curvature on the narrow neck occupies a smaller surface area.

We have derived the limit of stability $k_c(H_0)$ by considering the effects of a small-amplitude modulation, $|a| \ll 1$.  In addition, going beyond the linear regime we integrate the energy functional semi-numerically.  Elliptic integrals were used to integrate terms in (Equation \ref{eq:Hcurv}) over the sinusoidal surface shape where applicable and remaining nontractable terms were integrated numerically.  A shape amplitude $a_{min}$ minimizing the energy was found for a grid of values of $(k,H_0)$. The values are shown as the background shading in Figure \ref{fig:numerical_H0_g0}.
The region where linear limit of stability and numerical results disagree is to be interpreted as a region where the unmodulated cylinder is a metastable state.  In fact this metastability is driven by the effective surface tension term and is known for the classic Plateau-Rayleigh instability.  In the classic, surface-tension-dominated case, for a range of wavenumbers $k\geq1$, shape amplitude $|a|=0$ is a metastable local minimum of surface area $A(a)$, a large nonzero shape amplitude is the global minimum.  As previously described by Carter and Glaeser \cite{carter1987}, the system is unstable once nucleated with a sufficiently large shape fluctuation $a$; a new criterion for instability can be formulated in terms of both $k$ and $a$.

The semi-numerical investigation additionally indicates that, particularly near $H_0=\pm 1$, an intermediate shape amplitude $0<|a|<1$ may be the energetic minimum.  The analysis is restricted to a single wavelength at a time; the real system may be unstable against fluctuations of smaller wavelengths. In Appendix \ref{chap:pinchoff} we estimate whether a modulated shape will indeed be stable by examining the stability of the narrow neck against fluctuations of smaller wavelengths.  For spontaneous curvatures around $H_0=\pm 1$, a slightly modulated channel shape is in fact the stable equilibrium, despite the wavenumber $k<k_c$ being marked as unstable by the linear analysis.

\section{Orientational Order}
\subsection{Configurations of orientational order}
Below the isotropic-$n$-atic transition temperature the coefficient $\alpha$ is negative and the interfacial layer is orientationally ordered.  

In the azimuthal gradient term of Equation \ref{eq:hi},
\begin{equation}
\label{eq:gradient}
|D_\theta \Psi|^2 =  |\partial_\theta \Psi|^2 + \frac{2 n A_\theta}{r^2(a)} \mathfrak{ Im} ((\partial_\theta \Psi) \Psi^* )+ n^2 |A_\theta|^2 |\Psi|^2,
\end{equation}
we note that the cross-term can take negative values, suggesting that gradient energy can be decreased by orientational order whose direction rotates as it winds around the cylinder in the azimuthal direction.  The factor $g^{\theta\theta}= 1/g_{\theta\theta} = 1/r^2$, explicitly written in the middle term, is also implicitly present in tensor inner products $|X_i|^2 $ in the other two terms.  
Moreover, the equation suggests that a certain handedness of rotation is selected for.  This is a consequence of our choice to represent $n$-atic rotational order by Equation \ref{eq:psidef} rather than its complex conjugate field, or equivalently, to use the charge $n$ rather than $-n$ in the coupling.  In reality,  states that are solutions to either set of equations occur; the chiral symmetry is spontaneously broken when a vortex state develops.

In analogy with superconductors, we distinguish Type I and Type II behavior.  In Type I superconductors, the material transitions directly from the superconducting state which expels an applied magnetic field ($n$-atic order which expels Gaussian curvature) to a non-superconducting state penetrated by a magnetic field (isotropic state on a modulated surface).  In Type II superconductors, on the other hand, there is an intermediate vortex state, where the material is in the superconducting state almost everywhere but its phase (orientation of order) rotates around point defects, at which the magnetic field (Gaussian curvature) is concentrated.  It is not surprising that behavior analogous to superconductors emerges, as the description of $n$-atic order coupled to surface curvature is adapted from the Landau-Ginzburg equations for superconductors.  The analogy with the Abrikosov vortex state for spheres has been pointed out by Park et al. \cite{park1992}, studying spheres.  On near-spherical objects the range of available curvatures is dictated  by size and topology, while on our sinusoidal model geometry, a wider range of local curvatures is accessible.  The modulated cylindrical model system thus bears a closer, but still constrained, resemblance to superconductors, where arbitrary external magnetic fields may be applied. 

On the given closed surface, the $n$-atic field can undergo $N/n$ rotations as it winds around the cylinder once, with integer $N$.  Such a field is represented by the mode $\Psi = |\Psi| e^{i N \theta }$, with the linearly varying orientation minimizing gradient energy. Plugging this trial mode into Equation \ref{eq:gradient}, the local azimuthal gradient term is proportional to 
\begin{equation}
\label{eq:localdth}
|D_\theta \Psi|^2 = (N+n A_\theta)^2 \frac{|\Psi|^2}{r^2(a)}
\end{equation}
The first transition, when the energy can be minimized by selecting $N=1$ rather than $N=0$ rotations, occurs at the axial location $kz =  m \pi$, where the spin connection attains its maximal value, at wavenumber

\begin{equation}
\label{eq:ktypeII}
k(a) = \frac{\sqrt{2+ a^2}}{a \sqrt{-2 + 8n^2}}.
\end{equation}
In the case of $n=6$ the lower bound wavenumber for onset of Type II behavior is $\sqrt{3/286} \approx 0.102$, whereas in the case of $n=1$ it is $1/\sqrt{2} \approx 0.707$.  Analogous calculations for the transition from other values of $N$ to $N+1$ states suggest that for $1$-atic fields, modes with more than $N=1$ azimuthal rotations are never energetically advantageous, while for hexatic fields the spectrum extends to $N=6$.  In other words, on one period of the surface vector order has either $0$ or $4$ defects, while for hexatic order it may have any number $4N$, up to $4N = 24$, of defects.  The wavenumber of onset predicted here is a lower bound based on local energy balance at locations $kz = m \pi$.  This theoretical lower bound depends on $n$ but not on values of field parameters $\alpha<0$, $c$.  The true wavenumber of onset for the whole system will be increased by the interplay of several additional factors, including the energetic cost of defect cores, axial gradients, and the fact that the spin connection is less extreme at other locations;  it does depend on coefficients $c$ and $\alpha$. 

\begin{figure}
 \centering
        \includegraphics[width=\textwidth]{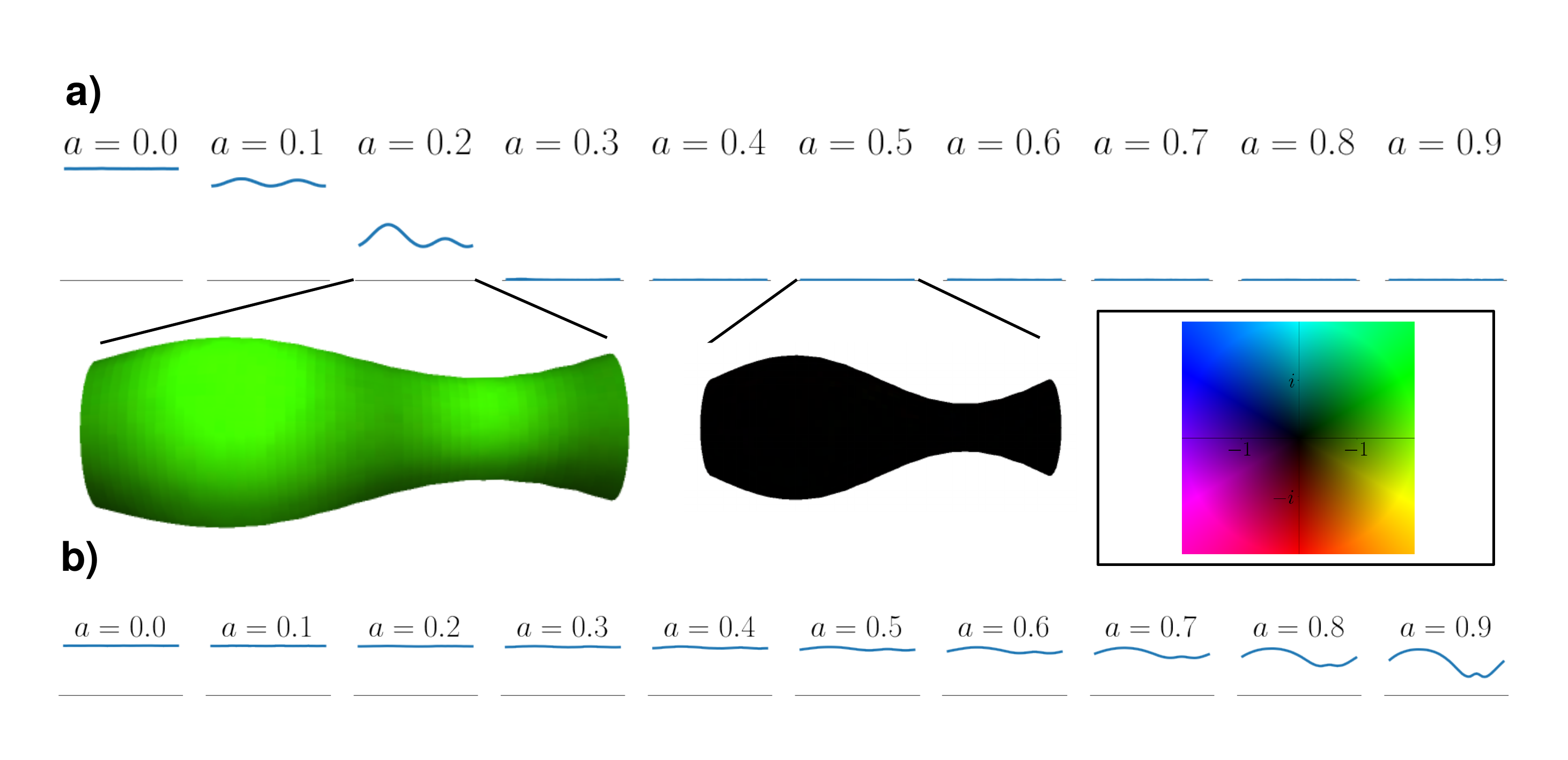}

\caption{Examples of Type I behaviour. Here the field configuration remains defect-free on a series of shape amplitudes from $a=0$ to $a=0.9$. The averaged field magnitude profile $\langle |\Psi| \rangle _{\theta,t}$ as a function of $z$ is shown as a graph, with a horizontal axis where $|\Psi|=0$ and $|\Psi|=2$ at $a=0$.  (a) Magnitude profiles from series of simulations with $n=6, \alpha=-4, c=6.5, k=0.9$. As shape amplitude increases, the field transitions from ordered field to locally depressed, then to everywhere isotropic. The characteristic magnitude profile is most depressed at two locations on either side of the narrow neck. Two simulation snapshots, a banded configuration and the uniform isotropic state, are rendered on the surface shape.  (b) Series of magnitude profiles from simulations $n=1, \alpha=-4, c=4.5, k=0.9$. The $1$-atic field is more weakly affected by shape modulations.  Inset: Colormap representing the complex-valued field.  Saturation, ranging from $0$ to $|\Psi|=2$, indicates field magnitude or amount of $n$-atic order, while hue indicates the phase of the complex field or the direction, modulo $1/n$, of orientational order.}
\label{fig:irrotationalseries}
\end{figure}

When the system is known to be confined to Type I behavior, the equations can be simplified: the field minimizing Equation $\ref{eq:hi}$ does not vary azimuthally, $\partial_\theta \Psi =0 $ and the gradient term is reduced to 
\begin{equation}
c|D_i \Psi|^2 =  c|\partial_z \Psi| + c n^2 |A_\theta|^2 |\Psi|^2.
\end{equation} 
The second term can be understood as an addition to coefficient $\alpha$, forming the axially varying effective coefficient $\alpha'(z) = \alpha + c n^2 |A_\theta|^2$.  The isotropic-$n$-atic transition temperature thus varies locally: it is increased on regions which are curved in the sense of having a nonzero spin connection.  The effect is largest in two regions adjacent to the narrowest location $kz = 3 \pi/2 $: $|A_\theta|^2$ is proportional to $\cos^2 kz $ but also to $1/r^2(z)$.  The resulting field configuration is one with constant phase and axially varying magnitude  $|\Psi|(z)$ minimizing the equation
\begin{equation}
\label{eq:HI1D}
\mathcal{H}_I^{N=0} = \int dS \left[ \alpha'(z) |\Psi|^2 +  c \partial_z |\Psi|\partial^z |\Psi|  +\frac{u}{2}|\Psi|^4 \right]
\end{equation}

We investigate field configurations on curved surface shapes using lattice-based Markov chain Monte Carlo simulations (Appendix \ref{chap:simulation}). Here, the order parameter field $\Psi(\bf{x})$ is represented as a discretized lattice of complex values.  Each simulation represents a system with material parameters $(n, \alpha, c)$ on a fixed surface shape $(k,a)$.  For a given field configuration, total field energy $\mathcal{H}_I[\Psi(\bf{x})]$ is calculated numerically as described in Appendix \ref{chap:simulation}.  Field configurations are evolved according to a low-temperature Monte Carlo sampling protocol so that the simulation converges on an energy-minimizing configuration.  For each surface shape and set of material parameters, we are then able to visually inspect the resulting field configuration as well as extract the minimized energy value.  

If, for a given set of material parameters and a wavenumber $k$, the field configuration remains vortex-free for all shape amplitudes $0\leq|a|<1$, we classify the material and shape as Type I.  Two examples in the Type I regime are shown in Figure \ref{fig:irrotationalseries}.  In this regime, increasing amplitude of shape modulations causes the field magnitude to decrease locally, resulting in modulated magnitude profiles $|\Psi|(z)$ minimizing the one-dimensionalized Equation \ref{eq:HI1D}. For large shape deformations the field is isotropic everywhere.

\begin{figure}
 \centering
        \includegraphics[width=\textwidth]{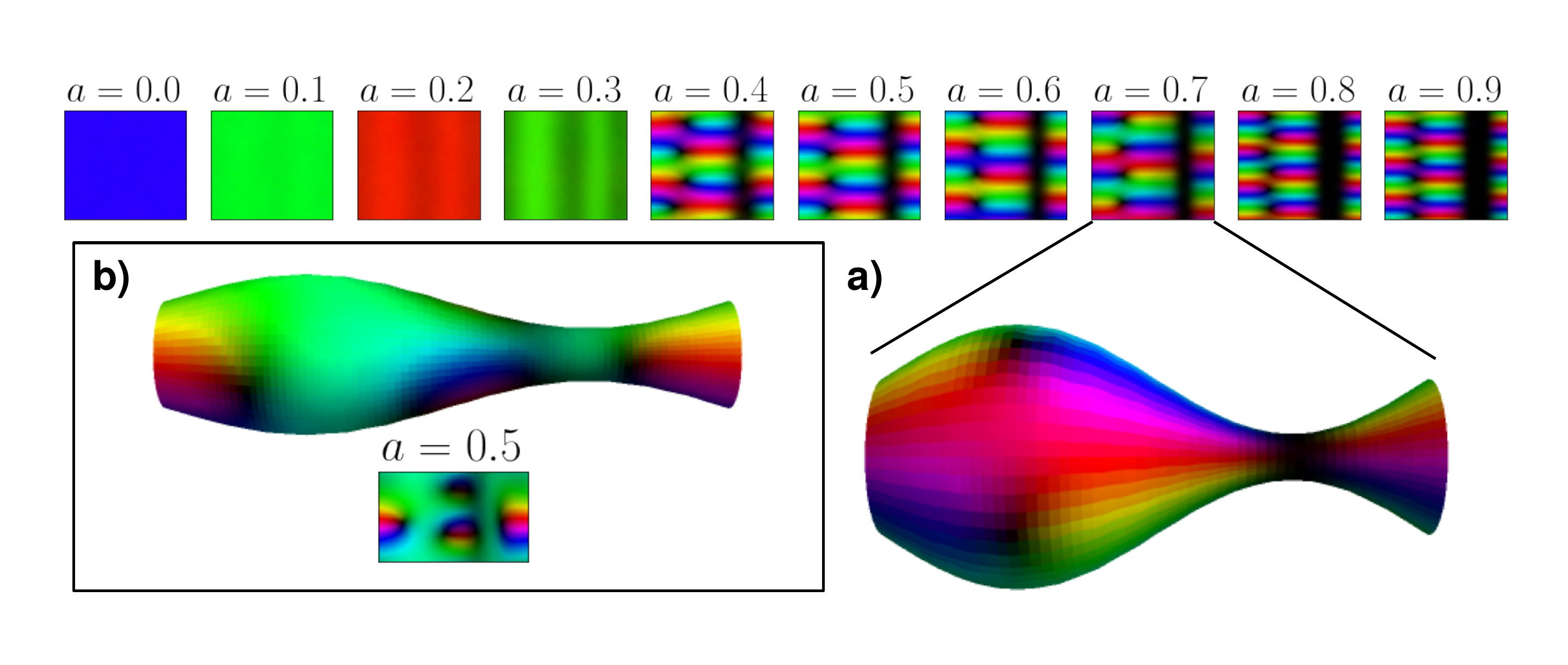}
\caption{Examples of Type II behavior in hexatic fields.  The final field configuration snapshot is shown on the $(z, \theta)$ plane using the complex colormap (Figure \ref{fig:irrotationalseries}, inset) to represent complex values. (a) Series in $a$  with $n=6, \alpha=-4$, $c=1.5$, $k=0.9$. As shape amplitude increases, the field here transitions from adapting to curvature via defect-free magnitude modulations to a state with $N=\pm2$ discrete azimuthal rotations and finally one with $N=\pm3$ rotations. There are $\Delta N = 2|N|$ visible defects on the widest part of the cylinder, while their counterparts on the narrow neck are often merged to an isotropic band.  (b) An example of a more complex banding pattern on a longer cylinder, simulation $n=6, \alpha=-4$, $c=2.5$, $k=0.6$, $a=0.5$.}
\label{fig:rotationalseries}
\end{figure}

On the other hand examples of field configurations from those simulations which displayed Type II behavior are shown in Figure \ref{fig:rotationalseries}.  As in the Type I cases, order is locally decreased, especially on either side of the narrow neck.  At higher shape amplitude $a$ we observe the predicted vortex state: between the widest region and the narrow neck orientational order rotates azimuthally $N$ times around the cylinder; on the other half it winds around the cylinder $N$ times in the opposite direction.  There must be $2N$ defects with the total defect charge $\pm 2N/n$ where the counterrotating bands meet at the narrowest and widest regions, for a total of $4N$ defects on the surface and a vanishing total defect charge.  The defects on the narrow neck are topologically present but are often visually obscured by an isotropic band.  At more extreme curvatures, the defects lie on a line at the narrowest/widest locations, whereas placement is more distributed in cases of less extreme curvature at lower wavenumber.  

We have not studied the system on the level of defects, rather defects appear as an emergent phenomenon. The analysis and simulation have not been directly informed by laws describing the interaction of defects and Gaussian curvature. It is therefore encouraging that the lattice simulation reproduces the well-known association between defect charge and local Gaussian curvature.  Seeing Gaussian curvature as an effective charge, we can apply the effective charge cancellation \cite{mesarec2016} principle, in which the sum of defect charge and Gaussian curvature charge in a small region $s$ approximates a locally neutral total charge:
\begin{equation}
\label{eq:defectcharge}
q M + \frac{1}{2\pi} \int_s dS K \approx 0, 
\end{equation}
where $M$ is the number of defects in a region $s$ and $q = \pm 1/n$ their charge.

Equation \ref{eq:defectcharge} can be derived in the special case of our cylinder-like system as follows:  As we reasoned above, at each axial location, in an idealized system the local number of azimuthal rotations $N$ takes the integer value most closely minimizing local gradient energy (Equation \ref{eq:localdth}), that is
\begin{equation}
\label{eq:18repeat}
N + n A_\theta \approx 0.
\end{equation}   On the cylinder-like shape, the total number and sign of defects in a region between two axial locations $z_1$ and $z_2$ is the sum of differences in local rotation numbers,  $\pm M = \int_{z_1}^{z_2} dz \sqrt{g_{zz}} \partial_z N$. We similarly take an axial derivative and integrate axially over the second term of Equation \ref{eq:18repeat} to obtain
\begin{equation}
\frac{\pm M}{n} + \int_{z_1}^{z_2} dz \sqrt{g_{zz}}  \partial_z A_\theta \approx 0.
\end{equation} By relating spin connection to Gaussian curvature via the Mermin-Ho theorem \cite{kamien2002} as $\partial_z A_\theta = \sqrt{g_{\theta \theta}} K$ and by labelling charge $q=\pm 1/n$, we recognize Equation \ref{eq:defectcharge}.  The factor of $2\pi$ is equivalent to additionally carrying out an azimuthal integration over the latter term.  

The number of defects on the wider half of the cylinder predicted via the effective topological charge cancellation mechanism roughly agrees that from our previous reasoning about rotation numbers $N$ at certain locations of maximal spin connection (Equation \ref{eq:ktypeII}) and with the number of defects appearing in simulation in the examples shown in Figure \ref{fig:rotationalseries}.  Like Equation \ref{eq:ktypeII}, the number of defects predicted by topological charge cancellation is an upper bound; for materials with large persistence length fewer defects may be realized.

\begin{figure}
 \centering
         \begin{subfigure}{.45\textwidth}
        \includegraphics[width=\linewidth]{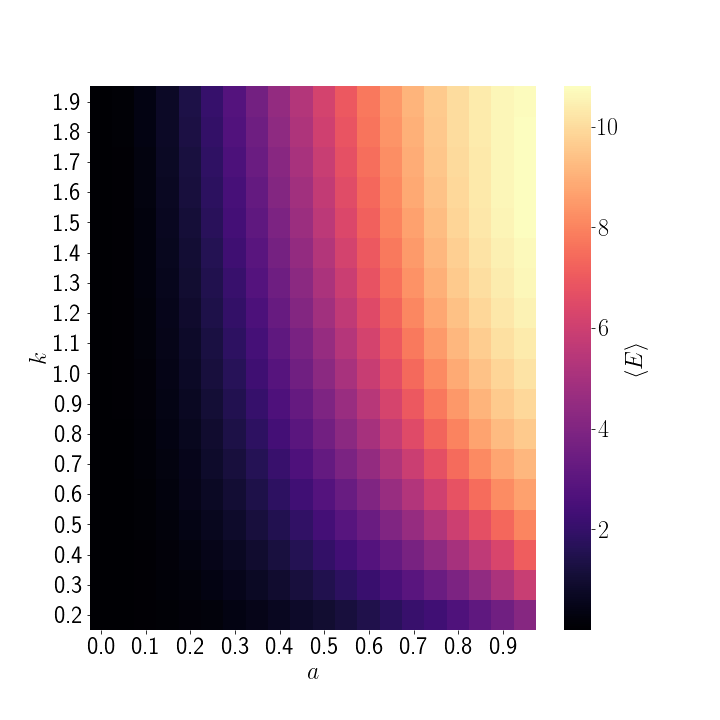}
        \subcaption{$n=1, \alpha=-1$ }
    \end{subfigure}
        \begin{subfigure}{.45\textwidth}
        \includegraphics[width=\linewidth]{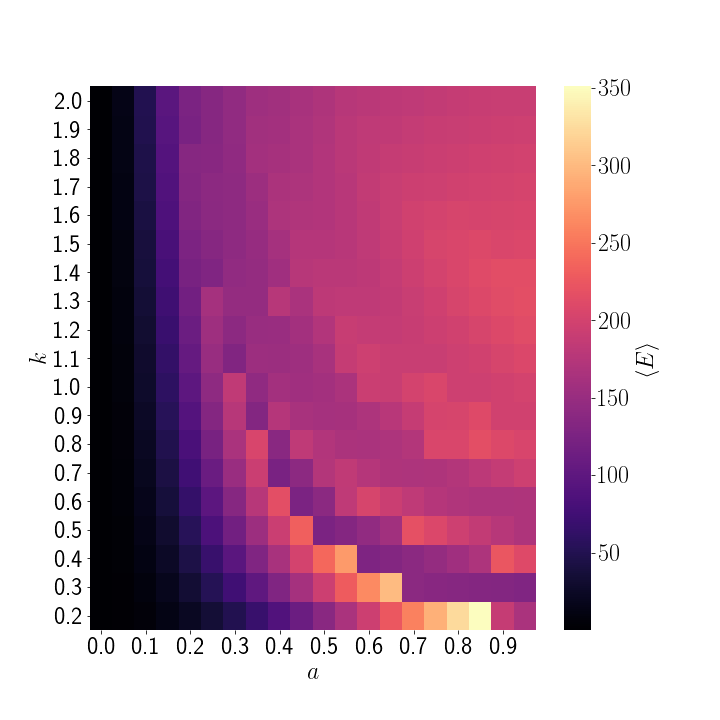}
        \subcaption{$n=6, \alpha=-4$}
    \end{subfigure}
\caption{Mean gradient energy $\langle E \rangle$ obtained from simulations on an array of surface shapes $(k,a)$.  (a) In the $n=1$ case field energy increases monotonically in $a$ for all $k$ examined here. (b) For $n=6$ there a multiple steps in the energy landscape, as the field transitions from the defect-free state to a spectrum of states $N$ with $4N$ defects.}
\label{fig:fieldenergy}
\end{figure}

For an array of simulations on a range of fixed surface shapes $(a,k)$ we collect average field energy $\langle E \rangle$ of the equilibrated simulations.  A vectorial and a hexatic example are shown in Figure \ref{fig:fieldenergy}.  In general, gradient energy density increases on surface shapes which are more curved in the sense of the spin connection, having larger shape amplitude and wavenumber.
For hexatic order, a stepped dependence of gradient energy on shape is apparent. The first discontinuity in energy corresponds to the onset of Type II behavior in the form of the first defect state with $N=1$, with each additional discontinuity corresponding to a transition to the next vortex state. The wavenumber of onset is about $k=0.2$ in the hexatic example with $c=1$.  In an analogous set of simulations with $n=1$ the onset of Type II behavior is not seen for any $k$ sampled here, up to $k=2.0$.  While a transition to Type II behavior is theoretically possible at some $k \geq 0.707$, for vector order it apparently occurs at larger wavenumber, smaller field stiffness $c$, or larger alignment $|\alpha|$ than those studied here.

\begin{figure}
 \centering
    \begin{subfigure}[b]{.45\textwidth}
        \includegraphics[width=\textwidth]{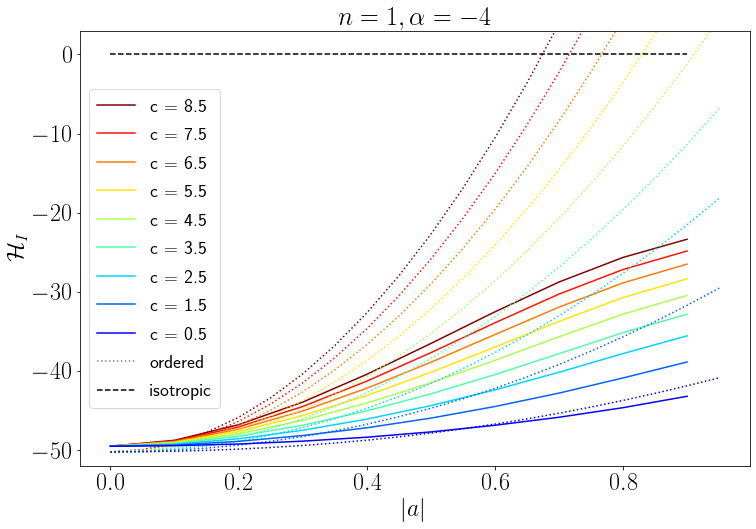}
        \subcaption{$n=1$}
    \end{subfigure}
        \begin{subfigure}[b]{.45\textwidth}
        \includegraphics[width=\textwidth]{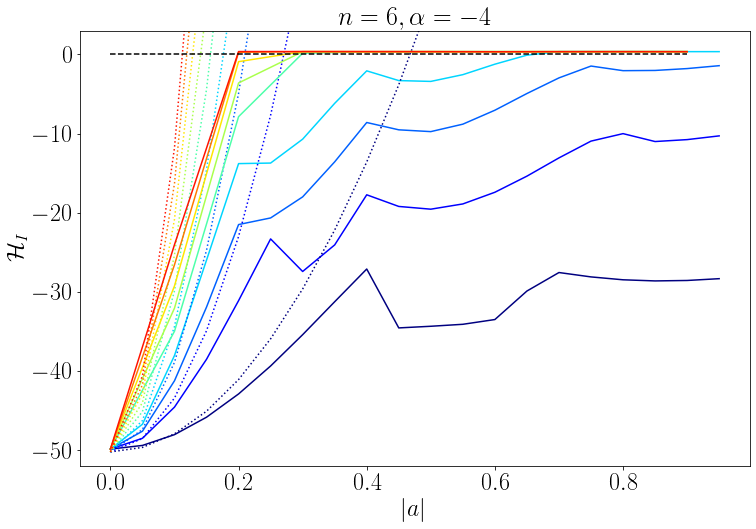}
        \subcaption{$n=6$}
    \end{subfigure}
\caption{Solid colored lines show mean field energy $\mathcal{H}_I(a)$ (per unit length of cylinder) retrieved from simulations on fixed surfaces with $k=0.9$ and a range of $a$.  As reference, we show the energy a uniformly ordered field would have on the curved surface shape (colored dotted lines), obtained by semi-numerically integrating Equation \ref{eq:uniformordered}, and the energy of a uniformly isotropic field (black dashed line, zero).  (a) For $n=1$, the field is ordered at low $a$ and at higher $a$ adjusts by locally adapting magnitude, lowering energy slightly below that of the perfectly uniform field on the same surface.  (b) For $n=6$ the field with high bending rigidity $c$ transitions to the uniformly isotropic state on curved surfaces.  For lower $c$ the hexatic field, in addition to locally adapting its magnitude, is able to decrease its energy by adopting rotational states with defects, resulting in a stepped energy response. }
\label{fig:energylandscapes}
\end{figure}

\subsection{Effect of orientational order on shape}
\begin{figure}
        \includegraphics[width=\textwidth]{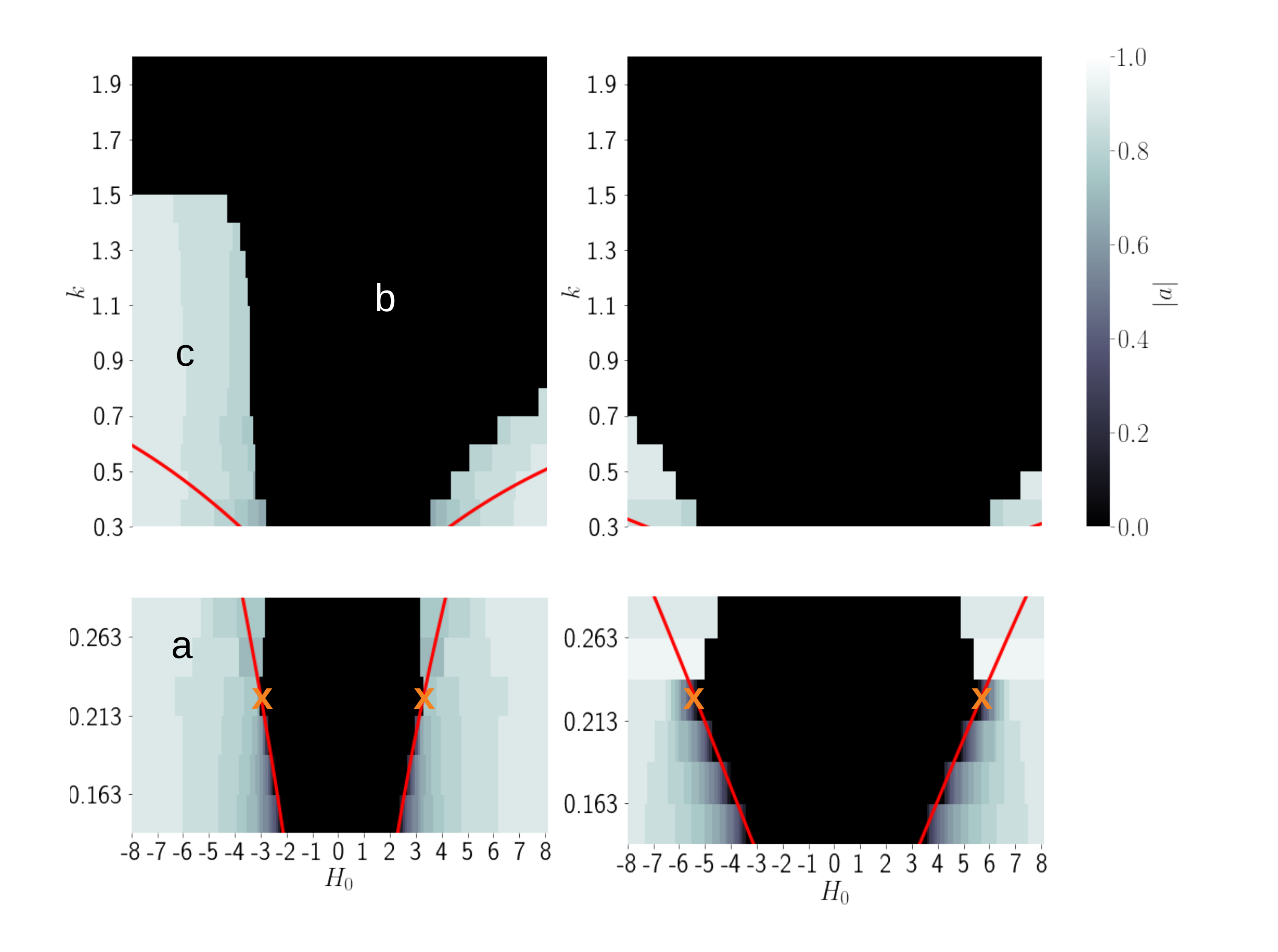}
\caption{The effect of hexatic order with $c=1$ and $\alpha=-1$ (left) or $\alpha=-4$ (right) on critical wavenumber as a function of spontaneous curvature.  The linear limit of stability is suppressed compared to the case with no order (compare Figure 2).  There is a qualitative transition from Type I field behavior at low wavenumbers $k_c \lesssim 0.215$, inducing a continuous shape transition adhering to the linear prediction, to Type II behavior at higher wavenumbers.  In the latter case the transition is discontinuous and the linear limit of stability gives an incomplete description of the system.  We mark the apparent critical endpoints (orange \texttt{x}) seperating continuous from discontinuous transitions.  As in Figure \ref{fig:numerical_H0_g0}, there are metastability regions (c) where linear analysis indicates perturbative stability of the unmodulated shape but numerics reveal that the global energy minimum of the system is a modulated shape. In metastable Type II systems, the metastability effect is enhanced by coinciding shape and vortex metastabilities and the ultimately stable state is a vortex configuration on a modulated shape.} 
\label{fig:fieldeffect}
\end{figure}

We now consider the effect of the interfacial order on the shape modulations.  First, in line with the previous linear stability analysis of a cylindrical interface with spontaneous curvature, we examine the linear effect of $n$-atic order.  Assuming the field is initially ordered ($\Psi(\bf{x}) = \Psi_0$) on the unperturbed cylinder, at small shape perturbations with amplitude $a \ll 1$, it can be shown that the induced changes in field configuration are negligible in terms of their energetic contribution.  To leading order the difference in internal energy is the gradient energy difference plus a term proportional to surface area change
\begin{equation}
\label{eq:chi}
\frac{\Delta \mathcal{H}}{A_0 a^2} = \chi k^2 - \frac{\alpha^2}{2u} \Delta A,
\end{equation} collecting field characteristics as $\chi=|\alpha|cn^2 /(2u)$. The second term, an energy difference proportional to change in surface area $\Delta A$, will be absorbed into surface tension.  Lenz and Nelson \cite{lenz2003}, treating a hexatic field with constant magnitude equivalent to $|\alpha|/u=1$, obtain a linear energy difference equivalent to the first term. Adding the effect of $n$-atic order to the energy difference Equation \ref{eq:diff} and again finding roots, the limit of stability is
\begin{equation}
\label{eq:kc_field}
\begin{aligned}
k_c(H_0, \chi) &= \frac{1}{2} \Big( 1-2 \gamma- 4 H_0 - 8 \chi \\&\pm \sqrt{8 (2 \gamma -3) + (-1+4 H_0 + 2 \gamma +8 \chi)^2}\Big)^{1/2}.
\end{aligned}
\end{equation}
with energy densities again in units where $\kappa=1$, $r_0=1$ and with $\gamma$ here representing both the surface energy density $f_0=-\alpha^2/(2 u)$ and an effective surface tension from spontaneous curvature: $\gamma = -\alpha^2/(2 u) +H_0^2/2$.
According to the linear analysis, a preferentially ordered field ($\alpha<0$) has a stabilizing effect on the system via the first term of equation \ref{eq:chi}.  While the second term, a negative effective surface tension from the ordered material, can theoretically induce an inverse Plateau-Rayleigh instability, where short-wavelength fluctuations grow to increase surface area, in the regime $|\alpha| \approx c $ studied here the stabilizing effect is dominant.  Examples of the effect of order on critical wavenumber are shown as red lines in Figure \ref{fig:fieldeffect}.  

As reference  energies, we calculate the energy a uniformly ordered field $\Psi(\bf{x})=\Psi_0$ and a uniformly isotropic field $\Psi(\bf{x})=0$ would have on a more heavily modulated surface.
The energy a spatially uniform ordered field $|\Psi|(\bf{x})= |\Psi_0|= \sqrt{-\alpha/u}$ would have on the modulated surface, including its contribution via $\gamma$, is
\begin{equation} 
\label{eq:uniformordered}
\mathcal{H}_I[\Psi(z)=\Psi_0]= 2\chi \int_S dS |A_\theta|^2 - \frac{\alpha^2}{2u}\int_S dS.
\end{equation}
The energy a uniformly isotropic field would have on the surface shape is 
\begin{equation}
\label{eq:uniformdisordered}
\mathcal{H}_I[\Psi(z)=0] = 0.
\end{equation}
For a range of shape amplitudes $a$,
Equation \ref{eq:uniformdisordered} was evaluated using elliptic integrals for the second term and numerical integration for the first term. 

The above analytic upper bounds describe spatially uniform field configurations. In our simulations, as in reality, the field representing orientational order is free to vary spatially.  The field was allowed to converge on energy-minimizing configurations in a series of simulations on fixed surfaces with increasing shape amplitudes. in Figure \ref{fig:energylandscapes}, we compare the resulting energy function $\mathcal{H}_I (a)$ from the series of simulations to the analytic reference energies. Unsurprisingly the energy of the simulated field is lower than that of either of the two uniform reference states.
In Figure \ref{fig:energylandscapes}-b, showing the $n=6$ case, stepped energy graphs are again the signature of Type II behavior.  Increasing shape amplitude leads to an increase in gradient energy, alleviated by the introduction of additional vortices.  The nonmonotonic dependence of field energy on shape amplitude implies that certain surface shapes $(k,a)$ are more compatible with the ordered interfacial layer than others, so that morphologies will be biased towards a discrete set of shapes.

Having gathered an array of field energy values $\mathcal{H}_I(k,a)$ from simulation, we combine these with semi-numerically integrated values $\mathcal{H}_E(k, a, H_0)$ from section \ref{chap:external}.  To predict the shape of the cylinder, for each point $(k,H_0)$ in parameter space we search numerically for the shape amplitude $a$ minimizing total system energy $\mathcal{H}(a) = \mathcal{H}_I(a)+\mathcal{H}_E(a)$.  The energy-minimizing shape amplitude, as a function of wavenumber and spontaneous curvature, is shown in Figure \ref{fig:fieldeffect} as the shaded background.  We compare the linear limit of stability predicted by Equation \ref{eq:kc_field}.   Below a certain wavenumber $k \approx 0.215$, where behavior is Type I, the linear limit of stability is a good description and the shape transition is continuous.
On the other hand at larger $k$, in the regime of Type II behavior, the transition from flat to modulated cylinders induced by increasing $|H_0|$ is discontinuous.  In this regime nonlinear effects, namely the emergence of the vortex state, become important.  The phenomenon coincides with and adds to the metastability due to nonlinear surface area changes which was already present in the classical Plateau-Rayleigh case; the metastable region is strongly extended and modified by effects of orientational order. 

\section{Discussion}
Cylinder-like vesicles and emulsion structures may be driven to a pearled morphology by factors such as a spontaneous curvature.  For a model system that is topologically simple but highly curved, we showcase significant nonlinear effects of local curvature on $n$-atic order and ultimately the effects of that order on the equilibrium morphology.

For shapes close to cylindrical, where we can expand linearly around an unmodulated cylindrical shape, $n$-atic order is trivially uniform and defect free. In the regime of relative high field stiffness studied here, it has a dominant stabilizing effect.  However, taking the possibility of externally induced modulated morphologies into account, strongly curvature-coupled order may interact with modulated surface shapes in complex ways.  First, on a surface shape with local Gaussian curvatures the isotropic-$n$-atic transition temperature of the ordered material is locally depressed, inducing a banded state of alternating ordered and isotropic regions.  Furthermore, at a 
threshold curvature, interfacial materials can adopt a vortex state, with orientation of order undergoing maximally $\pm N$ full rotations as the field winds around the cylinder azimuthally and with $4N$ defects.  By considering the maximal number of rotations that could be induced locally at the location of maximal spin connection, a lower bound, where the onset of the defect state becomes possible, is identified.  Simulations reveal the global emergence of a defect state; the complex interplay of a number of factors increases the wavenumber of onset above the lower bound estimate.

The delineation into systems which attain a defect state, analogous to a vortex state in Type II superconductor, from those that transition directly from ordered to isotropic fields, is significant for predicting the morphology of cylinders with both spontaneous curvature and order. In the latter case the transition is continuous and the linearized theory is a good description.  The system has a critical endpoint at a certain wavenumber and spontaneous curvature, above which nonlinear effects in the interplay between order and curvature strongly influence morphology.
Within the regime of excess defects, there are further transitions between discrete states with $4N$ defects, which will bias the spectrum of morphologies towards a discrete spectrum of wavelength-amplitude combinations.

In both cases, on certain regions which are more curved in the sense of the spin connection, the isotropic-$n$-atic transition temperature is effectively increased by curvature; order can be thought of as locally `melted' by curvature.  Both effects - the quasi-high-temperature phenomenon of decreased order, and the low-temperature state of isolated defects - coexist in the same system.  Working with the general formulation of Equation \ref{eq:hi}, rather than a representation as defects in a constant-magnitude field, which is well-suited for the low-temperature regime \cite{foltin2000}, is crucial to revealing this polymorphism. 
The resulting banded pattern of order and disorder is reminiscent of the banded partitioning of different species of lipid molecule, compatible with different spontaneous curvatures, on modulated cylinders.  Cases have been observed experimentally by Yanagisawa et al. \cite{yanagisawa2010} and extensively studied on fixed snowman surfaces by Rinaldin et al \cite{rinaldin2020}. Among other differences the field describing lipid composition has $Z_2$ rather than $O_n$ symmetry and will not form defects. Interestingly it nevertheless has some features in common with the $n$-atic field in its Type I regime.

The system has been assumed to obey a Hamiltonian with certain rotational symmetries.  i) Mechanical bending rigidities in Equation \ref{eq:hh} are isotropic.    In several interesting biological systems, such as cell membranes with curvature-inducing proteins, curvature elasticity can be strongly anisotropic \cite{iglic2005}.  ii) In our model the order parameter field is not coupled to any extrinsic or mean curvature terms in Equation \ref{eq:hi}.  Such a coupling exists more or less prominently in various systems, from interactions through the bulk phase between hexatically arranged spheres \cite{law2020} to the prominently mean-curvature-inducing properties of ordered domains of inclusions \cite{kralj-iglic2000}.  The focus on extrinsic curvature coupling has allowed us to study a field theory with unbroken continuous $O_n$ symmetry and the associated vortex state.  The research could be extended to systems where the symmetry is broken by additional curvature coupling terms.  We expect intermediate behaviour, with a weakened or absent vortex state.

The axisymmetric model is incomplete with respect to the shape deformations examined.  It is clear that in a similar experimental system as for cooled emulsion droplets, as in spherical droplets faceting effects will become important where there are defects.  A more precise study of the relevant shape variations, beyond the scope of the model surfaces examined here, warrants further study.

Representing order of particles and flexible interfaces simultaneously in simulation is an ongoing challenge.  With the continuum field representation of orientational order, general principles can be explored.  A low-temperature lattice-based Monte Carlo simulation, adapted to include multiple effects of an underlying curved surfaces, was used to obtain field configurations in the low-temperature limit. Due to peculiarities of statistical field theories represented on a non-uniform lattice, further modifications are needed before our simulation can accurately represent fluctuations and extend our exploration to the regime of high-temperature sampling.  The model and simulation protocol used here allow an efficient exploration of parameter space: taking advantage of the linearly additive formulation of effects of internal and external energies, the model allows combination in post-processing.  Simulation results verify and extend the analytical predictions presented here and, crucially, allow us to delineate regions of parameter space where linearized and quasi-one-dimensional descriptions of the system are sufficient from those where nonlinear effects dominate.

\begin{acknowledgments}
The authors thank Patrick Warren for useful discussion.  This work was undertaken on ARC3, part of the High Performance Computing facilities at the University of Leeds, UK.  JK acknowledges, with thanks, funding from the EPSRC through the SOFI (Soft Matter and Functional Interfaces) Centre for Doctoral Training (grant EP/L015536/1).  
\end{acknowledgments}

\appendix

\section{Self-similar stability analysis}
\label{chap:pinchoff}
We study cylinders with a single sinusoidal modulation of wavenumber $k$, which may be seen as the longest wavelength a system can support.  Instability against fluctuations of this wavelength or any smaller wavelengths indicate instability of the system.

Furthermore, where stabilizing curvature-related effects are present, initial growth of a perturbation does not necessarily indicate continuing exponential growth and breakup into spheres.  The numerical study indicates a shape amplitude $a$ minimizing the system energy.
When the value is sufficiently close to $|a|=1$ (corresponding to infinitesimally thin necks), we may assume the model of a sinusoidal cylinder breaks down and the system will break up into spheres.  However, in some cases, small nonzero shape amplitudes, corresponding to a slightly modulated shape, are indicates as the energetic minimum.

We estimate whether this is truly a stable configuration, or whether smaller perturbations on the resulting narrow neck will develop further and cause pinchoff, by recursively considering the narrow neck as a self-similar subsystem.

\begin{figure}
\centering
\includegraphics[width=.45\textwidth]{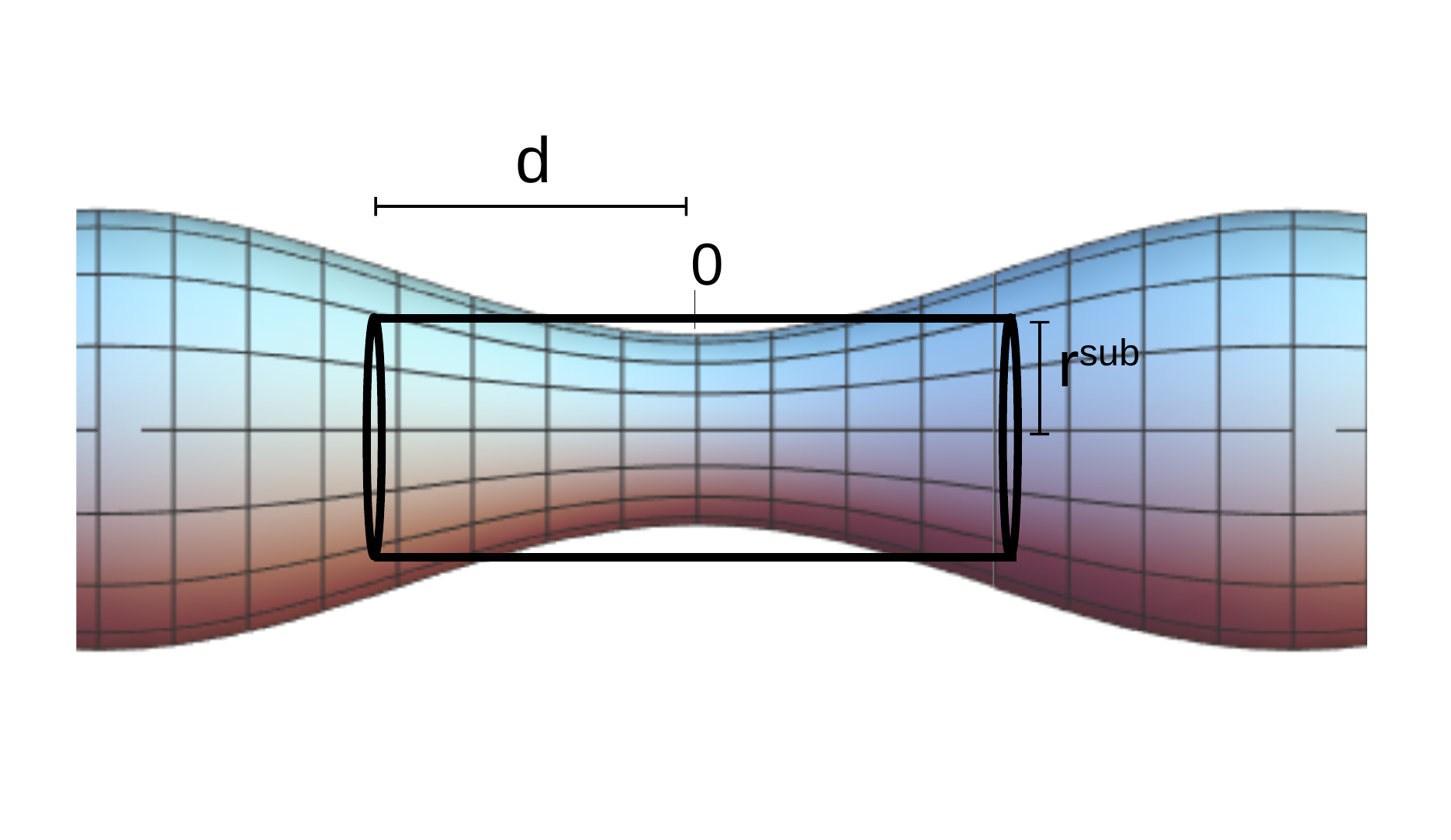}
\caption{To estimate stability of a modulated morphology against shorter-wavelength fluctuations, we approximate a subsystem centered on the narrow neck as an initially straight cylinder.  The subsystem has length $2d$ and radius $r^{sub}=\langle r \rangle$.}
\label{fig:subsystem}
\end{figure}

We consider a subsystem of length $2 d$, where $d$ may range from $0$ to $\pi/2$, centered on the narrowest location (Figure \ref{fig:subsystem}).
While the true radius of the narrow neck varies axially, 
we approximate the subsystem as a cylinder of uniform radius by taking the mean radius  

\begin{equation}
\begin{aligned}
\langle r \rangle &= \frac{ \int_0^d r(a)\left( 1-  a \cos( kz)\right) dz}{d}\\
& =  r(a)(1-a/d \sin(kz))
\end{aligned}
\end{equation}
as the subsystem radius $r^{sub}$.
In this approximation,  initial curvature of the subsystem is also neglected.  The pre-existing axial curvature may slightly stabilize the subsystem relative to our estimate.
Finally the narrow neck is assumed to obey the same energetics as the larger system, including a volume constraint.  

We find the the critical wavenumber $k_c^{sub}(d)$ of a subsystem of length $2d$ by applying Equation \ref{eq:kc} to the self-similar subsystem.  By absorbing factors of $r^{sub}$ into $\kappa$, the subsystem has a larger effective bending rigidity $\kappa' = \kappa / (r^{sub})^2$.  Because surface tension is given in units of $\kappa'$, it has a smaller relative surface tension $\gamma^{sub} = \gamma (r^{sub})^2$.   Effective spontaneous curvature is, in units of the subsystem radius,  to $H'_0 = H_0 r^{sub}$.  

Starting from a system with $k$, $H_0$, and the shape amplitude $a$ indicated by the numerical analysis in section \ref{chap:external}, we check numerically whether, for any $d \in (0, \pi/2)$, there exists a subsystem whose length is larger than its critical wavenumber $2\pi / k_c^{sub}(d)$.

\begin{figure}
\centering
\includegraphics[width=.45\textwidth]{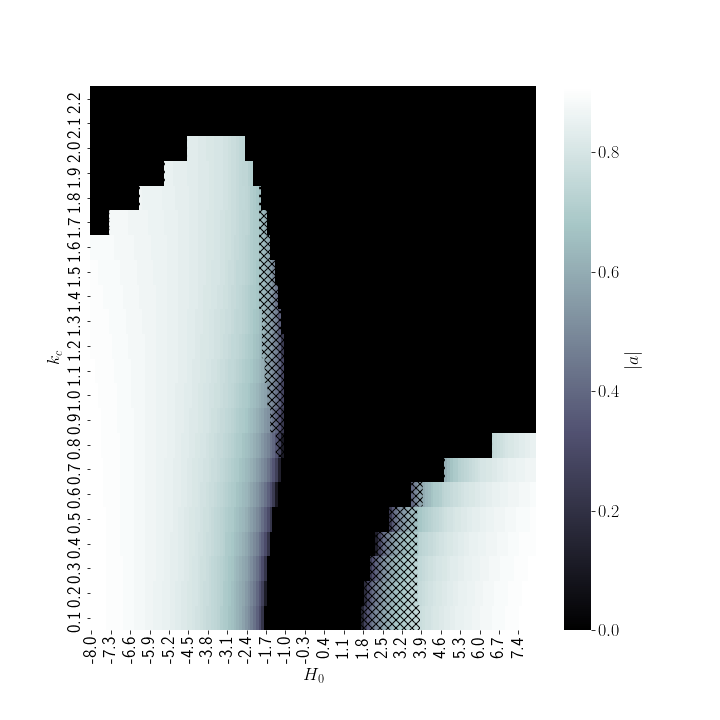}
\caption{The hatched area indicates systems where, according to our self-similar estimate, the modulated shape is linearly stable against all smaller-wavelength fluctuations.}
\label{fig:pinchoff}
\end{figure}

Where such a linearly unstable subsystem exists, this does not necessarily indicate the modulated cylinder will breakup.  The linearly unstable narrow neck may develop a large-amplitude fluctuation which leads to pinch-off, or the subsystem may itself have an energetic minimum at a mildly modulated shape, which does not itself develop any further instabilities.  We leave the further development unanswered and merely confirm that there exist mildly modulated channel shapes for which no linearly unstable subsystem exists.  For at least some cases, indicated by the hatched area on Figure \ref{fig:pinchoff},  the mildly modulated channel shape indicated by the numerical analysis is indeed a stable equilibrium.

\section{Simulation}
\label{chap:simulation}
We turn to stochastic  simulation to find field configurations minimizing Equation \ref{eq:hi} on modulated surface shapes, and ultimately to estimate the configurations ($a, \Psi(\bf{x})$) jointly minimizing total energy of the system modulated with a fixed wavenumber $k$.  
Our simulation is a lattice-based Markov chain Monte Carlo simulation.  
While the simulation is capable of simultaneously sampling field configurations $\Psi(\bf{x})$ and surface shape amplitude $a$, we find that the parameter space is more efficiently covered by collecting a database of field energy on an array of fixed surface shapes $(k,a)$.  This data from simulation is then combined with the external energy $\mathcal{H}_E$, which can be quickly calculated, \textit{post hoc}.  

The $n$-atic order parameter field is represented as a two-dimensional lattice of $N = (50 \times \lfloor50/k\rfloor)$ complex values $\Psi(z_i, \theta_j)$.  On the unperturbed cylinder each lattice cell represents an area with dimensions  $l^0_\theta = 2\pi/50 $, $\ell_z^0 = 2 \pi / \lfloor50/k\rfloor$, while on modulated cylinders cell areas are $\ell_\theta = \ell_\theta^0 \sqrt{g_{\theta \theta}}$, $\ell_z = \ell_z^0 \sqrt{g_{zz}}$.
Each lattice cell is associated with a values $g_{zz}$ and $g_{\theta \theta}$ of the metric as well as a value of the spin connection $A_\theta$. The values are retrieved according to Equations \ref{eq:g} and \ref{eq:connection} based on the axial location $z$ of the cell and shape amplitude $a$ of the surface.

The simulation is intended to discover field configurations in the low-temperature limit of the theory, where fluctuations do not play a role.  To nevertheless allow adequate thermal Monte Carlo sampling for convergence on the minimizing field configuration, after nondimensionalizing, temperature was set to $T=0.001$ throughout the study.
As the basic update of the simulation, a single lattice cell is selected and its value $\Psi(z_i, \theta_j)$ is updated by an increment drawn from a complex Gaussian distribution.  The new value is accepted or rejected with Boltzmann probability $P=e^{-\frac{\Delta E}{k_B T}}$ according to the usual Metropolis algorithm.  

The energy difference on changing the value at a single lattice cell is calculated as
\begin{equation}
\begin{aligned}
\label{eq:totalsimenergy}
\frac{\Delta E_{i,j}}{\ell_z \ell_\theta} &= E^{mag}(\Psi^{p}_{i,j}) -  E^{mag}(\Psi^{i}_{i,j}) \\&+ E^{zgrad}(\Psi^{p}_{i,j}, \Psi^{i}_{i-1, j}) - E^{zgrad}(\Psi^{i}_{i, j}, \Psi^{i}_{i-1, j}) 
\\&+ E^{zgrad}(\Psi^{i}_{i+1, j}, \Psi^{p}_{i, j}) - E^{zgrad}(\Psi^{i}_{i+1, j}, \Psi^{i}_{i, j}) 
\\&+ E^{\theta grad}(\Psi^{p}_{i, j-1}, \Psi^{i}_{i, j}) - E^{\theta grad}(\Psi^{i}_{i,j-1}, \Psi^{i}_{i,j},) 
\\&+ E^{\theta grad}(\Psi^{i}_{i,j+1}, \Psi^{p}_{i, j}) - E^{\theta grad}(\Psi^{i}_{i,j}, \Psi^{i}_{i,j+1}),
\end{aligned}
\end{equation}
where superscript $i$ marks initial values and $p$ the proposed new value.  Note the energy difference $\Delta E_{i,j}$ used in the Metropolis algorithm is scaled by cell area $\ell_z \ell_\theta$.
The magnitude-associated part of the energy is
\begin{equation}
\begin{aligned}
E^{mag}(\Psi_{i,j}) &=(\alpha+c n^2 |A_\theta|^2)|\Psi_{i,j}|^2+ \frac{u}{2} |\Psi_{i,j}|^4.
\end{aligned}
\end{equation}

For the gradient terms, as a numerical implementation of the derivative operator, a simple backwards derivative is chosen: 
\begin{equation}
\begin{aligned}
\partial_z \Psi_{i,j} \approx \frac{\Psi(z_i, \theta_j) - \Psi(z_{i-1}, \theta_j) }{\ell_z}\\
\partial_\theta \Psi_{i,j} \approx \frac{\Psi(z_i, \theta_j) - \Psi(z_{i}, \theta_{j-1})} {\ell_\theta}.\\
\end{aligned}
\end{equation}
Consequently a change to the value at lattice site $(z_i, \theta_j)$ also affects the gradient energy ascribed to neighboring lattice sites $(z_{i+1}, \theta_j)$ and $(z_i, \theta_{j+1})$.  For this reason change in gradient terms at both the site itself and at two adjacent site appears in Equation \ref{eq:totalsimenergy}.  Gradient energy ascribed to a site $(z_i,\theta_j)$ is 

\begin{equation}
E^{z grad}(\Psi_{i,j}, \Psi_{i-1,j})  = cn^2\left|\frac {\Psi_{i,j}- \Psi_{i-1, j}}{\ell_z}\right|^2
\end{equation}
and
\begin{equation}
\begin{aligned}
E^{\theta grad}&(\Psi_{i,j}, \Psi_{i, j-1}) = cn^2\left| \frac{\Psi_{i,j} - \Psi_{i, j-1}}{\ell_\theta}\right|^2 \\&+ 
2 cn \mathfrak{Im} \left[ \left(A_\theta \Psi_{i,j} \right)^* \left( \frac{\Psi_{i,j} - \Psi_{i, j-1}}{\ell_\theta}\right) \right].
\end{aligned}
\end{equation}


To form one simulations step, $100 \times N$ randomly chosen cells are updated. Simulations were usually run for $50 000$ steps.  Wavenumbers $k \leq0.8$ in Figure \ref{fig:fieldeffect} were instead run for $200 000$ steps and for the additional data in the same figure, covering wavenumbers $k=0.15$ to $0.3$, the simulation was run for $100 000$ steps as the larger lattices are slower to reach equilibration.

An adaptive sampling protocol was used.  Proposed values are drawn from a complex Gaussian distribution centered on the old value and having sampling width $\sigma=\sigma_l \sqrt{g}$.   The base sampling width $\sigma_l$ is evolved according to the adaptive algorithm recomended by Garthwaite et al. \cite{garthwaite2016}.  The target acceptance rate was set to $0.5$.  Rather than maintaining a separate sampling width for each lattice cell, the same base sampling width $\sigma_l$ is used, scaled in proportion to area, for each cell.  

After each step, the total field energy $E$ is evaluated and recorded as the sum
\begin{equation}
\begin{aligned}
E(t) = \sum_{ij} \Big[ &E^{mag}(\Psi_{i,j}) + E^{zgrad}(\Psi_{i,j}, \Psi_{i-1, j}) \\&+ E^{\theta grad}(\Psi_{i, j}, \Psi_{i, j-1}) \Big] 
\end{aligned}
\end{equation}
over lattice cells $i,j$.  The energy includes an energy contribution from thermal fluctuations, but in these low-temperature simulations it is negligible: by equipartition theorem it is approximately $0.4$ energy units per unit length of cylinder.

\begin{figure}
\centering
\begin{subfigure}{.45\textwidth}
\includegraphics[width=\textwidth]{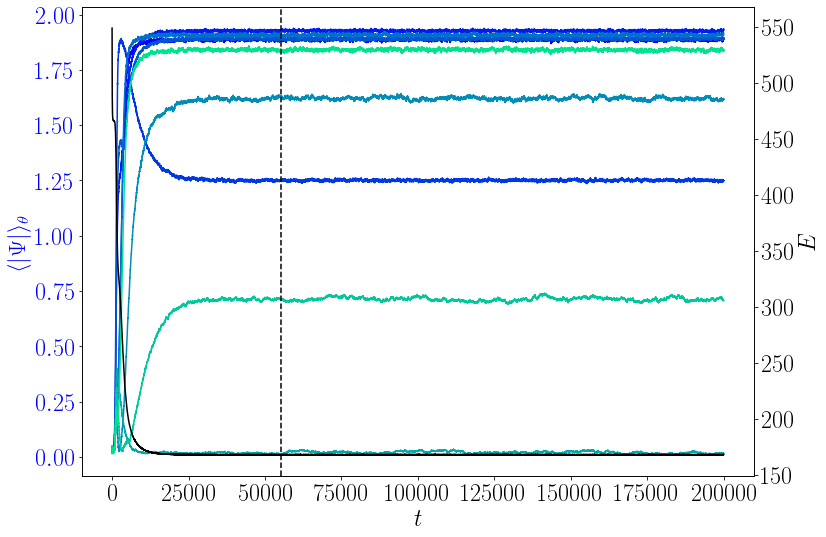}
\subcaption{Simulation $n=6, \alpha= -4, c=1, k=0.6, a=0.9$. A well-equilibrated but non-trivial Type II example with cutoff point $\tau_0=55080$. }
\end{subfigure}
\begin{subfigure}{.45\textwidth}
\includegraphics[width=\textwidth]{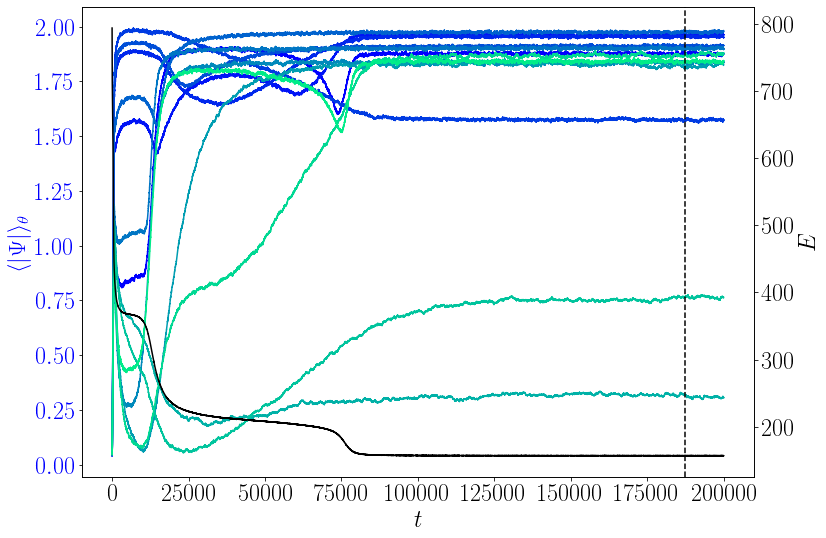}
\subcaption{Simulation $n=6, \alpha= -4, c=1, k=0.4, a=0.8$. An example with a late cutoff point $\tau_0=187440$.}
\end{subfigure}
\caption{Time series in total field energy $E$ and additional observables $\langle |\Psi(z_i)| \rangle_{\theta}$, recorded at every 10th axial location $z_i$, over simulation time.  Early trends in these values correspond to the emergence and axial migration of defects.}
\label{fig:equilibration}
\end{figure}

In addition to total field energy $E$, after each step an azimuthally averaged magnitude $\langle |\Psi(z_i)| \rangle_{\theta}$ is recorded for every 10th axial location $z_i$.  The system has several continuous symmetries; the global phase of the field and azimuthal placement of defects may vary slowly over simulation time without indicating non-equilibration. On the other hand the quantity $\langle |\Psi(z_i)| \rangle_{\theta}$ is a suitable observable for detecting equilibration.  
The optimal production dataset was detected by the method of minimizing statistical inefficiency \cite{chodera2016}, as implemented in the pymbar timeseries module \cite{chodera2007,shirts2008}.  Autocorrelation in the timeseries of each observable was analysed independently and the maximum cutoff time from among these was used as a global cutoff for the simulation.  By inspection of time series (Figure \ref{fig:equilibration}), it corresponds well to equilibration.   For $n=6$ simulations, cutoff times were on average $\bar{\tau}_0=11140$ for the set of simulations of length $50000$, $\bar{\tau}_0=79160$ for simulations of length $200000$, and $\bar{\tau}_0=16804$ for simulations of length $100000$. For the $n=1$ simulations, which all remained in the relatively trivial defect-free states, simulation length was always $50000$, and the average cutoff time was $\bar{\tau}_0 = 10203$. Mean quantities $\mathcal{H}_I = \langle E \rangle_t$ reported above refer to averages over the production region of the simulation, from cutoff $\tau_0$ to the end.

The lattice simulation scheme on curved surfaces as used here contains uncorrected artefacts related to varying cutoff lengthscales and the representation of fluctuations.  
Unusually, in this simulation scheme the surface is covered by a sometimes extremely distorted lattice.
On the more curved surface shapes, we have simulation cells representing highly differing areas in the same simulations as well as highly anisotropic cell dimensions.  The problem becomes apparent when considering that $\Psi_{ij}$ represents a spatially averaged value of the orientational order of molecules or particles within a given lattice cell.  Realistically, magnitude and variance of this value should differ depending on the area included in the spatial averaging.  The naive implementation does not correctly reflect the scale dependence of the field theory subject to thermal fluctuations.
For more accurate high-temperature simulations, coefficients in Equation \ref{eq:hi} can be renormalized to better represent the behavior of the same field theory at the different lengthscales present in the same simulation.  The naive scaling of energies by cell area used here is the correct renormalization for a Gaussian field theory and the first term in the exact expression for rescaled coefficients of the full field theory.
Without higher-order corrections to coefficients, our simulation protocol is exact only in the low-temperature limit.  On the other hand, the uncorrected implementation used here has the advantage of conceptual and computational simplicity. While a balance must be struck between avoiding higher sampling temperatures and achieving adequate Monte Carlo sampling, the simulation protocol performs well approaching the low-temperature limit and is suitable to give qualitative results.  

\bibliographystyle{apsrev4-1}
\bibliography{paper1bib}

\begin{thebibliography}{34}%
\makeatletter
\providecommand \@ifxundefined [1]{%
 \@ifx{#1\undefined}
}%
\providecommand \@ifnum [1]{%
 \ifnum #1\expandafter \@firstoftwo
 \else \expandafter \@secondoftwo
 \fi
}%
\providecommand \@ifx [1]{%
 \ifx #1\expandafter \@firstoftwo
 \else \expandafter \@secondoftwo
 \fi
}%
\providecommand \natexlab [1]{#1}%
\providecommand \enquote  [1]{``#1''}%
\providecommand \bibnamefont  [1]{#1}%
\providecommand \bibfnamefont [1]{#1}%
\providecommand \citenamefont [1]{#1}%
\providecommand \href@noop [0]{\@secondoftwo}%
\providecommand \href [0]{\begingroup \@sanitize@url \@href}%
\providecommand \@href[1]{\@@startlink{#1}\@@href}%
\providecommand \@@href[1]{\endgroup#1\@@endlink}%
\providecommand \@sanitize@url [0]{\catcode `\\12\catcode `\$12\catcode
  `\&12\catcode `\#12\catcode `\^12\catcode `\_12\catcode `\%12\relax}%
\providecommand \@@startlink[1]{}%
\providecommand \@@endlink[0]{}%
\providecommand \url  [0]{\begingroup\@sanitize@url \@url }%
\providecommand \@url [1]{\endgroup\@href {#1}{\urlprefix }}%
\providecommand \urlprefix  [0]{URL }%
\providecommand \Eprint [0]{\href }%
\providecommand \doibase [0]{http://dx.doi.org/}%
\providecommand \selectlanguage [0]{\@gobble}%
\providecommand \bibinfo  [0]{\@secondoftwo}%
\providecommand \bibfield  [0]{\@secondoftwo}%
\providecommand \translation [1]{[#1]}%
\providecommand \BibitemOpen [0]{}%
\providecommand \bibitemStop [0]{}%
\providecommand \bibitemNoStop [0]{.\EOS\space}%
\providecommand \EOS [0]{\spacefactor3000\relax}%
\providecommand \BibitemShut  [1]{\csname bibitem#1\endcsname}%
\let\auto@bib@innerbib\@empty
\bibitem [{\citenamefont {Binks}(2002)}]{binks2002}%
  \BibitemOpen
  \bibfield  {author} {\bibinfo {author} {\bibfnamefont {B.~P.}\ \bibnamefont
  {Binks}},\ }\href@noop {} {\bibfield  {journal} {\bibinfo  {journal} {Curr
  Opin Colloid Interface Sci}\ }\textbf {\bibinfo {volume} {7}},\ \bibinfo
  {pages} {21} (\bibinfo {year} {2002})}\BibitemShut {NoStop}%
\bibitem [{\citenamefont {Marin}\ \emph {et~al.}(2020)\citenamefont {Marin},
  \citenamefont {Tkachev}, \citenamefont {Sloutskin},\ and\ \citenamefont
  {Deutsch}}]{marin2020}%
  \BibitemOpen
  \bibfield  {author} {\bibinfo {author} {\bibfnamefont {O.}~\bibnamefont
  {Marin}}, \bibinfo {author} {\bibfnamefont {M.}~\bibnamefont {Tkachev}},
  \bibinfo {author} {\bibfnamefont {E.}~\bibnamefont {Sloutskin}}, \ and\
  \bibinfo {author} {\bibfnamefont {M.}~\bibnamefont {Deutsch}},\ }\href@noop
  {} {\bibfield  {journal} {\bibinfo  {journal} {Curr Opin Colloid Interface
  Sci}\ } (\bibinfo {year} {2020})}\BibitemShut {NoStop}%
\bibitem [{\citenamefont {Guttman}\ \emph {et~al.}(2016)\citenamefont
  {Guttman}, \citenamefont {Sapir}, \citenamefont {Schultz}, \citenamefont
  {Butenko}, \citenamefont {Ocko}, \citenamefont {Deutsch},\ and\ \citenamefont
  {Sloutskin}}]{guttman2016}%
  \BibitemOpen
  \bibfield  {author} {\bibinfo {author} {\bibfnamefont {S.}~\bibnamefont
  {Guttman}}, \bibinfo {author} {\bibfnamefont {Z.}~\bibnamefont {Sapir}},
  \bibinfo {author} {\bibfnamefont {M.}~\bibnamefont {Schultz}}, \bibinfo
  {author} {\bibfnamefont {A.~V.}\ \bibnamefont {Butenko}}, \bibinfo {author}
  {\bibfnamefont {B.~M.}\ \bibnamefont {Ocko}}, \bibinfo {author}
  {\bibfnamefont {M.}~\bibnamefont {Deutsch}}, \ and\ \bibinfo {author}
  {\bibfnamefont {E.}~\bibnamefont {Sloutskin}},\ }\href@noop {} {\bibfield
  {journal} {\bibinfo  {journal} {Proc National Acad Sci U S A}\ }\textbf
  {\bibinfo {volume} {113}},\ \bibinfo {pages} {493} (\bibinfo {year}
  {2016})}\BibitemShut {NoStop}%
\bibitem [{\citenamefont {Garc{\'\i}a-Aguilar}\ \emph
  {et~al.}(2021)\citenamefont {Garc{\'\i}a-Aguilar}, \citenamefont {Fonda},
  \citenamefont {Sloutskin},\ and\ \citenamefont {Giomi}}]{garcia-aguilar2021}%
  \BibitemOpen
  \bibfield  {author} {\bibinfo {author} {\bibfnamefont {I.}~\bibnamefont
  {Garc{\'\i}a-Aguilar}}, \bibinfo {author} {\bibfnamefont {P.}~\bibnamefont
  {Fonda}}, \bibinfo {author} {\bibfnamefont {E.}~\bibnamefont {Sloutskin}}, \
  and\ \bibinfo {author} {\bibfnamefont {L.}~\bibnamefont {Giomi}},\
  }\href@noop {} {\bibfield  {journal} {\bibinfo  {journal} {Phys Rev Lett}\
  }\textbf {\bibinfo {volume} {126}},\ \bibinfo {pages} {038001} (\bibinfo
  {year} {2021})}\BibitemShut {NoStop}%
\bibitem [{\citenamefont {Abkarian}\ \emph {et~al.}(2007)\citenamefont
  {Abkarian}, \citenamefont {Subramaniam}, \citenamefont {Kim}, \citenamefont
  {Larsen}, \citenamefont {Yang},\ and\ \citenamefont {Stone}}]{abkarian2007}%
  \BibitemOpen
  \bibfield  {author} {\bibinfo {author} {\bibfnamefont {M.}~\bibnamefont
  {Abkarian}}, \bibinfo {author} {\bibfnamefont {A.~B.}\ \bibnamefont
  {Subramaniam}}, \bibinfo {author} {\bibfnamefont {S.-H.}\ \bibnamefont
  {Kim}}, \bibinfo {author} {\bibfnamefont {R.~J.}\ \bibnamefont {Larsen}},
  \bibinfo {author} {\bibfnamefont {S.-M.}\ \bibnamefont {Yang}}, \ and\
  \bibinfo {author} {\bibfnamefont {H.~A.}\ \bibnamefont {Stone}},\ }\href@noop
  {} {\bibfield  {journal} {\bibinfo  {journal} {Phys Rev Lett}\ }\textbf
  {\bibinfo {volume} {99}},\ \bibinfo {pages} {188301} (\bibinfo {year}
  {2007})}\BibitemShut {NoStop}%
\bibitem [{\citenamefont {Lenz}\ and\ \citenamefont {Nelson}(2003)}]{lenz2003}%
  \BibitemOpen
  \bibfield  {author} {\bibinfo {author} {\bibfnamefont {P.}~\bibnamefont
  {Lenz}}\ and\ \bibinfo {author} {\bibfnamefont {D.~R.}\ \bibnamefont
  {Nelson}},\ }\href {\doibase 10.1103/PhysRevE.67.031502} {\bibfield
  {journal} {\bibinfo  {journal} {Phys Rev E}\ }\textbf {\bibinfo {volume}
  {67}},\ \bibinfo {pages} {19} (\bibinfo {year} {2003})}\BibitemShut {NoStop}%
\bibitem [{\citenamefont {Evans}\ \emph {et~al.}(1996)\citenamefont {Evans},
  \citenamefont {Bowman}, \citenamefont {Leung}, \citenamefont {Needham},\ and\
  \citenamefont {Tirrell}}]{eevans1996}%
  \BibitemOpen
  \bibfield  {author} {\bibinfo {author} {\bibfnamefont {E.}~\bibnamefont
  {Evans}}, \bibinfo {author} {\bibfnamefont {H.}~\bibnamefont {Bowman}},
  \bibinfo {author} {\bibfnamefont {A.}~\bibnamefont {Leung}}, \bibinfo
  {author} {\bibfnamefont {D.}~\bibnamefont {Needham}}, \ and\ \bibinfo
  {author} {\bibfnamefont {D.}~\bibnamefont {Tirrell}},\ }\href {\doibase
  10.1126/science.273.5277.933} {\bibfield  {journal} {\bibinfo  {journal}
  {Science}\ }\textbf {\bibinfo {volume} {273}},\ \bibinfo {pages} {933}
  (\bibinfo {year} {1996})}\BibitemShut {NoStop}%
\bibitem [{\citenamefont {Bar-Ziv}\ and\ \citenamefont
  {Moses}(1994)}]{bar-ziv1994}%
  \BibitemOpen
  \bibfield  {author} {\bibinfo {author} {\bibfnamefont {R.}~\bibnamefont
  {Bar-Ziv}}\ and\ \bibinfo {author} {\bibfnamefont {E.}~\bibnamefont
  {Moses}},\ }\href@noop {} {\bibfield  {journal} {\bibinfo  {journal} {Phys
  Rev Lett}\ }\textbf {\bibinfo {volume} {73}},\ \bibinfo {pages} {1392}
  (\bibinfo {year} {1994})}\BibitemShut {NoStop}%
\bibitem [{\citenamefont {Stone}(1994)}]{stone1994}%
  \BibitemOpen
  \bibfield  {author} {\bibinfo {author} {\bibfnamefont {H.~A.}\ \bibnamefont
  {Stone}},\ }\href@noop {} {\bibfield  {journal} {\bibinfo  {journal} {Annu
  Rev Fluid Mech}\ }\textbf {\bibinfo {volume} {26}},\ \bibinfo {pages} {65}
  (\bibinfo {year} {1994})}\BibitemShut {NoStop}%
\bibitem [{\citenamefont {Li}\ \emph {et~al.}(2019)\citenamefont {Li},
  \citenamefont {Klebes}, \citenamefont {Dobnikar},\ and\ \citenamefont
  {Clegg}}]{li2019}%
  \BibitemOpen
  \bibfield  {author} {\bibinfo {author} {\bibfnamefont {T.}~\bibnamefont
  {Li}}, \bibinfo {author} {\bibfnamefont {J.}~\bibnamefont {Klebes}}, \bibinfo
  {author} {\bibfnamefont {J.}~\bibnamefont {Dobnikar}}, \ and\ \bibinfo
  {author} {\bibfnamefont {P.~S.}\ \bibnamefont {Clegg}},\ }\href@noop {}
  {\bibfield  {journal} {\bibinfo  {journal} {Chem Comm}\ }\textbf {\bibinfo
  {volume} {55}},\ \bibinfo {pages} {5575} (\bibinfo {year}
  {2019})}\BibitemShut {NoStop}%
\bibitem [{\citenamefont {Evans}(1995)}]{evans1995}%
  \BibitemOpen
  \bibfield  {author} {\bibinfo {author} {\bibfnamefont {R.~M.~L.}\
  \bibnamefont {Evans}},\ }\href {\doibase 10.1051/jp2:1995147} {\bibfield
  {journal} {\bibinfo  {journal} {Eur Phys J}\ }\textbf {\bibinfo {volume}
  {5}},\ \bibinfo {pages} {507} (\bibinfo {year} {1995})},\ \Eprint
  {http://arxiv.org/abs/9410010} {arXiv:9410010 [cond-mat]} \BibitemShut
  {NoStop}%
\bibitem [{\citenamefont {Bowick}\ and\ \citenamefont
  {Giomi}(2009)}]{bowick2009}%
  \BibitemOpen
  \bibfield  {author} {\bibinfo {author} {\bibfnamefont {M.~J.}\ \bibnamefont
  {Bowick}}\ and\ \bibinfo {author} {\bibfnamefont {L.}~\bibnamefont {Giomi}},\
  }\href@noop {} {\bibfield  {journal} {\bibinfo  {journal} {Adv Phys}\
  }\textbf {\bibinfo {volume} {58}},\ \bibinfo {pages} {449} (\bibinfo {year}
  {2009})}\BibitemShut {NoStop}%
\bibitem [{\citenamefont {Granek}(1996)}]{granek1996}%
  \BibitemOpen
  \bibfield  {author} {\bibinfo {author} {\bibfnamefont {R.}~\bibnamefont
  {Granek}},\ }\href@noop {} {\bibfield  {journal} {\bibinfo  {journal}
  {Langmuir}\ }\textbf {\bibinfo {volume} {12}},\ \bibinfo {pages} {5022}
  (\bibinfo {year} {1996})}\BibitemShut {NoStop}%
\bibitem [{\citenamefont {Cha{\"{i}}eb}\ and\ \citenamefont
  {Rica}(1998)}]{chaieb1998}%
  \BibitemOpen
  \bibfield  {author} {\bibinfo {author} {\bibfnamefont {S.}~\bibnamefont
  {Cha{\"{i}}eb}}\ and\ \bibinfo {author} {\bibfnamefont {S.}~\bibnamefont
  {Rica}},\ }\href {\doibase 10.1103/PhysRevE.58.7733} {\bibfield  {journal}
  {\bibinfo  {journal} {Phys Rev E}\ }\textbf {\bibinfo {volume} {58}},\
  \bibinfo {pages} {7733} (\bibinfo {year} {1998})}\BibitemShut {NoStop}%
\bibitem [{\citenamefont {Mesarec}\ \emph {et~al.}(2017)\citenamefont
  {Mesarec}, \citenamefont {Kurioz}, \citenamefont {Igli{\v{c}}}, \citenamefont
  {G{\'o}{\'z}d{\'z}},\ and\ \citenamefont {Kralj}}]{mesarec2017}%
  \BibitemOpen
  \bibfield  {author} {\bibinfo {author} {\bibfnamefont {L.}~\bibnamefont
  {Mesarec}}, \bibinfo {author} {\bibfnamefont {P.}~\bibnamefont {Kurioz}},
  \bibinfo {author} {\bibfnamefont {A.}~\bibnamefont {Igli{\v{c}}}}, \bibinfo
  {author} {\bibfnamefont {W.}~\bibnamefont {G{\'o}{\'z}d{\'z}}}, \ and\
  \bibinfo {author} {\bibfnamefont {S.}~\bibnamefont {Kralj}},\ }\href@noop {}
  {\bibfield  {journal} {\bibinfo  {journal} {Crystals}\ }\textbf {\bibinfo
  {volume} {7}},\ \bibinfo {pages} {153} (\bibinfo {year} {2017})}\BibitemShut
  {NoStop}%
\bibitem [{\citenamefont {Helfrich}(1973)}]{helfrich1973}%
  \BibitemOpen
  \bibfield  {author} {\bibinfo {author} {\bibfnamefont {W.}~\bibnamefont
  {Helfrich}},\ }\href {\doibase 10.1515/znc-1973-11-1209} {\bibfield
  {journal} {\bibinfo  {journal} {Zeitschrift f{\"{u}}r Naturforschung C}\
  }\textbf {\bibinfo {volume} {28}},\ \bibinfo {pages} {693} (\bibinfo {year}
  {1973})}\BibitemShut {NoStop}%
\bibitem [{\citenamefont {Kamien}(2002)}]{kamien2002}%
  \BibitemOpen
  \bibfield  {author} {\bibinfo {author} {\bibfnamefont {R.~D.}\ \bibnamefont
  {Kamien}},\ }\href {\doibase https://doi.org/10.1103/RevModPhys.74.953}
  {\bibfield  {journal} {\bibinfo  {journal} {Reviews of Modern Physics}\
  }\textbf {\bibinfo {volume} {74}},\ \bibinfo {pages} {953} (\bibinfo {year}
  {2002})}\BibitemShut {NoStop}%
\bibitem [{\citenamefont {Park}\ \emph {et~al.}(1992)\citenamefont {Park},
  \citenamefont {Lubensky},\ and\ \citenamefont {MacKintosh}}]{park1992}%
  \BibitemOpen
  \bibfield  {author} {\bibinfo {author} {\bibfnamefont {J.-M.}\ \bibnamefont
  {Park}}, \bibinfo {author} {\bibfnamefont {T.~C.}\ \bibnamefont {Lubensky}},
  \ and\ \bibinfo {author} {\bibfnamefont {F.~C.}\ \bibnamefont {MacKintosh}},\
  }\href {\doibase 10.1209/0295-5075/20/3/015} {\bibfield  {journal} {\bibinfo
  {journal} {EPL}\ }\textbf {\bibinfo {volume} {20}},\ \bibinfo {pages} {279}
  (\bibinfo {year} {1992})},\ \Eprint {http://arxiv.org/abs/9606105}
  {arXiv:9606105 [cond-mat]} \BibitemShut {NoStop}%
\bibitem [{\citenamefont {Evans}(1996)}]{evans1996}%
  \BibitemOpen
  \bibfield  {author} {\bibinfo {author} {\bibfnamefont {R.~M.~L.}\
  \bibnamefont {Evans}},\ }\href@noop {} {\bibfield  {journal} {\bibinfo
  {journal} {Phys Rev E}\ }\textbf {\bibinfo {volume} {53}},\ \bibinfo {pages}
  {935} (\bibinfo {year} {1996})}\BibitemShut {NoStop}%
\bibitem [{\citenamefont {{Kumar Alageshan}}\ \emph {et~al.}(2017)\citenamefont
  {{Kumar Alageshan}}, \citenamefont {Chakrabarti},\ and\ \citenamefont
  {Hatwalne}}]{kumaralageshan2017}%
  \BibitemOpen
  \bibfield  {author} {\bibinfo {author} {\bibfnamefont {J.}~\bibnamefont
  {{Kumar Alageshan}}}, \bibinfo {author} {\bibfnamefont {B.}~\bibnamefont
  {Chakrabarti}}, \ and\ \bibinfo {author} {\bibfnamefont {Y.}~\bibnamefont
  {Hatwalne}},\ }\href {\doibase 10.1103/PhysRevE.95.042806} {\bibfield
  {journal} {\bibinfo  {journal} {Physical Review E}\ }\textbf {\bibinfo
  {volume} {95}},\ \bibinfo {pages} {042806} (\bibinfo {year}
  {2017})}\BibitemShut {NoStop}%
\bibitem [{\citenamefont {Foltin}\ and\ \citenamefont
  {Lehrer}(2000)}]{foltin2000}%
  \BibitemOpen
  \bibfield  {author} {\bibinfo {author} {\bibfnamefont {G.}~\bibnamefont
  {Foltin}}\ and\ \bibinfo {author} {\bibfnamefont {R.~A.}\ \bibnamefont
  {Lehrer}},\ }\href {\doibase 10.1088/0305-4470/33/6/304} {\bibfield
  {journal} {\bibinfo  {journal} {J Phys A}\ }\textbf {\bibinfo {volume}
  {33}},\ \bibinfo {pages} {1139} (\bibinfo {year} {2000})},\ \Eprint
  {http://arxiv.org/abs/9901095} {9901095 [cond-mat]} \BibitemShut {NoStop}%
\bibitem [{\citenamefont {Chaikin}\ \emph {et~al.}(1995)\citenamefont
  {Chaikin}, \citenamefont {Lubensky},\ and\ \citenamefont
  {Witten}}]{chaikin1995}%
  \BibitemOpen
  \bibfield  {author} {\bibinfo {author} {\bibfnamefont {P.~M.}\ \bibnamefont
  {Chaikin}}, \bibinfo {author} {\bibfnamefont {T.~C.}\ \bibnamefont
  {Lubensky}}, \ and\ \bibinfo {author} {\bibfnamefont {T.~A.}\ \bibnamefont
  {Witten}},\ }\href@noop {} {\emph {\bibinfo {title} {Principles of condensed
  matter physics}}},\ Vol.~\bibinfo {volume} {10}\ (\bibinfo  {publisher}
  {Cambridge university press Cambridge},\ \bibinfo {year} {1995})\BibitemShut
  {NoStop}%
\bibitem [{\citenamefont {Boedec}\ \emph {et~al.}(2014)\citenamefont {Boedec},
  \citenamefont {Jaeger},\ and\ \citenamefont {Leonetti}}]{boedec2014}%
  \BibitemOpen
  \bibfield  {author} {\bibinfo {author} {\bibfnamefont {G.}~\bibnamefont
  {Boedec}}, \bibinfo {author} {\bibfnamefont {M.}~\bibnamefont {Jaeger}}, \
  and\ \bibinfo {author} {\bibfnamefont {M.}~\bibnamefont {Leonetti}},\ }\href
  {\doibase 10.1017/jfm.2014.34} {\bibfield  {journal} {\bibinfo  {journal}
  {Journal of Fluid Mechanics}\ }\textbf {\bibinfo {volume} {743}},\ \bibinfo
  {pages} {262} (\bibinfo {year} {2014})}\BibitemShut {NoStop}%
\bibitem [{\citenamefont {Carter}\ and\ \citenamefont
  {Glaeser}(1987)}]{carter1987}%
  \BibitemOpen
  \bibfield  {author} {\bibinfo {author} {\bibfnamefont {W.}~\bibnamefont
  {Carter}}\ and\ \bibinfo {author} {\bibfnamefont {A.}~\bibnamefont
  {Glaeser}},\ }\href@noop {} {\bibfield  {journal} {\bibinfo  {journal} {Mater
  Sci and Eng}\ }\textbf {\bibinfo {volume} {89}},\ \bibinfo {pages} {L41}
  (\bibinfo {year} {1987})}\BibitemShut {NoStop}%
\bibitem [{\citenamefont {Mesarec}\ \emph {et~al.}(2016)\citenamefont
  {Mesarec}, \citenamefont {G{\'o}{\'z}d{\'z}}, \citenamefont {Igli{\v{c}}},\
  and\ \citenamefont {Kralj}}]{mesarec2016}%
  \BibitemOpen
  \bibfield  {author} {\bibinfo {author} {\bibfnamefont {L.}~\bibnamefont
  {Mesarec}}, \bibinfo {author} {\bibfnamefont {W.}~\bibnamefont
  {G{\'o}{\'z}d{\'z}}}, \bibinfo {author} {\bibfnamefont {A.}~\bibnamefont
  {Igli{\v{c}}}}, \ and\ \bibinfo {author} {\bibfnamefont {S.}~\bibnamefont
  {Kralj}},\ }\href@noop {} {\bibfield  {journal} {\bibinfo  {journal} {Sci
  Rep}\ }\textbf {\bibinfo {volume} {6}},\ \bibinfo {pages} {1} (\bibinfo
  {year} {2016})}\BibitemShut {NoStop}%
\bibitem [{\citenamefont {Yanagisawa}\ \emph {et~al.}(2010)\citenamefont
  {Yanagisawa}, \citenamefont {Imai},\ and\ \citenamefont
  {Taniguchi}}]{yanagisawa2010}%
  \BibitemOpen
  \bibfield  {author} {\bibinfo {author} {\bibfnamefont {M.}~\bibnamefont
  {Yanagisawa}}, \bibinfo {author} {\bibfnamefont {M.}~\bibnamefont {Imai}}, \
  and\ \bibinfo {author} {\bibfnamefont {T.}~\bibnamefont {Taniguchi}},\
  }\href@noop {} {\bibfield  {journal} {\bibinfo  {journal} {Phys Rev E}\
  }\textbf {\bibinfo {volume} {82}} (\bibinfo {year} {2010})}\BibitemShut
  {NoStop}%
\bibitem [{\citenamefont {Rinaldin}\ \emph {et~al.}(2020)\citenamefont
  {Rinaldin}, \citenamefont {Fonda}, \citenamefont {Giomi},\ and\ \citenamefont
  {Kraft}}]{rinaldin2020}%
  \BibitemOpen
  \bibfield  {author} {\bibinfo {author} {\bibfnamefont {M.}~\bibnamefont
  {Rinaldin}}, \bibinfo {author} {\bibfnamefont {P.}~\bibnamefont {Fonda}},
  \bibinfo {author} {\bibfnamefont {L.}~\bibnamefont {Giomi}}, \ and\ \bibinfo
  {author} {\bibfnamefont {D.~J.}\ \bibnamefont {Kraft}},\ }\href {\doibase
  10.1038/s41467-020-17432-w} {\bibfield  {journal} {\bibinfo  {journal} {Nat
  Commun}\ }\textbf {\bibinfo {volume} {11}},\ \bibinfo {pages} {4314}
  (\bibinfo {year} {2020})},\ \Eprint {http://arxiv.org/abs/1804.08596}
  {arXiv:1804.08596} \BibitemShut {NoStop}%
\bibitem [{\citenamefont {Igli{\v{c}}}\ \emph {et~al.}(2005)\citenamefont
  {Igli{\v{c}}}, \citenamefont {Babnik}, \citenamefont {Gimsa},\ and\
  \citenamefont {Kralj-Igli{\v{c}}}}]{iglic2005}%
  \BibitemOpen
  \bibfield  {author} {\bibinfo {author} {\bibfnamefont {A.}~\bibnamefont
  {Igli{\v{c}}}}, \bibinfo {author} {\bibfnamefont {B.}~\bibnamefont {Babnik}},
  \bibinfo {author} {\bibfnamefont {U.}~\bibnamefont {Gimsa}}, \ and\ \bibinfo
  {author} {\bibfnamefont {V.}~\bibnamefont {Kralj-Igli{\v{c}}}},\ }\href@noop
  {} {\bibfield  {journal} {\bibinfo  {journal} {J Phys A}\ }\textbf {\bibinfo
  {volume} {38}},\ \bibinfo {pages} {8527} (\bibinfo {year}
  {2005})}\BibitemShut {NoStop}%
\bibitem [{\citenamefont {Law}\ \emph {et~al.}(2020)\citenamefont {Law},
  \citenamefont {Dean}, \citenamefont {Miller},\ and\ \citenamefont
  {Kusumaatmaja}}]{law2020}%
  \BibitemOpen
  \bibfield  {author} {\bibinfo {author} {\bibfnamefont {J.~O.}\ \bibnamefont
  {Law}}, \bibinfo {author} {\bibfnamefont {J.~M.}\ \bibnamefont {Dean}},
  \bibinfo {author} {\bibfnamefont {M.~A.}\ \bibnamefont {Miller}}, \ and\
  \bibinfo {author} {\bibfnamefont {H.}~\bibnamefont {Kusumaatmaja}},\
  }\href@noop {} {\bibfield  {journal} {\bibinfo  {journal} {Soft Matter}\
  }\textbf {\bibinfo {volume} {16}},\ \bibinfo {pages} {8069} (\bibinfo {year}
  {2020})}\BibitemShut {NoStop}%
\bibitem [{\citenamefont {Kralj-Igli{\v{c}}}\ \emph {et~al.}(2000)\citenamefont
  {Kralj-Igli{\v{c}}}, \citenamefont {Igli{\v{c}}}, \citenamefont
  {H{\"a}gerstrand},\ and\ \citenamefont {Peterlin}}]{kralj-iglic2000}%
  \BibitemOpen
  \bibfield  {author} {\bibinfo {author} {\bibfnamefont {V.}~\bibnamefont
  {Kralj-Igli{\v{c}}}}, \bibinfo {author} {\bibfnamefont {A.}~\bibnamefont
  {Igli{\v{c}}}}, \bibinfo {author} {\bibfnamefont {H.}~\bibnamefont
  {H{\"a}gerstrand}}, \ and\ \bibinfo {author} {\bibfnamefont {P.}~\bibnamefont
  {Peterlin}},\ }\href@noop {} {\bibfield  {journal} {\bibinfo  {journal} {Phys
  Rev E}\ }\textbf {\bibinfo {volume} {61}},\ \bibinfo {pages} {4230} (\bibinfo
  {year} {2000})}\BibitemShut {NoStop}%
\bibitem [{\citenamefont {Garthwaite}\ \emph {et~al.}(2016)\citenamefont
  {Garthwaite}, \citenamefont {Fan},\ and\ \citenamefont
  {Sisson}}]{garthwaite2016}%
  \BibitemOpen
  \bibfield  {author} {\bibinfo {author} {\bibfnamefont {P.~H.}\ \bibnamefont
  {Garthwaite}}, \bibinfo {author} {\bibfnamefont {Y.}~\bibnamefont {Fan}}, \
  and\ \bibinfo {author} {\bibfnamefont {S.~A.}\ \bibnamefont {Sisson}},\
  }\href {\doibase 10.1080/03610926.2014.936562} {\bibfield  {journal}
  {\bibinfo  {journal} {Commun Stat Theory Methods}\ }\textbf {\bibinfo
  {volume} {45}},\ \bibinfo {pages} {5098} (\bibinfo {year} {2016})},\ \Eprint
  {http://arxiv.org/abs/1006.3690} {arXiv:1006.3690} \BibitemShut {NoStop}%
\bibitem [{\citenamefont {Chodera}(2016)}]{chodera2016}%
  \BibitemOpen
  \bibfield  {author} {\bibinfo {author} {\bibfnamefont {J.~D.}\ \bibnamefont
  {Chodera}},\ }\href@noop {} {\bibfield  {journal} {\bibinfo  {journal} {J
  Chem Theory Comput}\ }\textbf {\bibinfo {volume} {12}},\ \bibinfo {pages}
  {1799} (\bibinfo {year} {2016})}\BibitemShut {NoStop}%
\bibitem [{\citenamefont {Chodera}\ \emph {et~al.}(2007)\citenamefont
  {Chodera}, \citenamefont {Swope}, \citenamefont {Pitera}, \citenamefont
  {Seok},\ and\ \citenamefont {Dill}}]{chodera2007}%
  \BibitemOpen
  \bibfield  {author} {\bibinfo {author} {\bibfnamefont {J.~D.}\ \bibnamefont
  {Chodera}}, \bibinfo {author} {\bibfnamefont {W.~C.}\ \bibnamefont {Swope}},
  \bibinfo {author} {\bibfnamefont {J.~W.}\ \bibnamefont {Pitera}}, \bibinfo
  {author} {\bibfnamefont {C.}~\bibnamefont {Seok}}, \ and\ \bibinfo {author}
  {\bibfnamefont {K.~A.}\ \bibnamefont {Dill}},\ }\href@noop {} {\bibfield
  {journal} {\bibinfo  {journal} {J Chem Theory Comput}\ }\textbf {\bibinfo
  {volume} {3}},\ \bibinfo {pages} {26} (\bibinfo {year} {2007})}\BibitemShut
  {NoStop}%
\bibitem [{\citenamefont {Shirts}\ and\ \citenamefont
  {Chodera}(2008)}]{shirts2008}%
  \BibitemOpen
  \bibfield  {author} {\bibinfo {author} {\bibfnamefont {M.~R.}\ \bibnamefont
  {Shirts}}\ and\ \bibinfo {author} {\bibfnamefont {J.~D.}\ \bibnamefont
  {Chodera}},\ }\href@noop {} {\bibfield  {journal} {\bibinfo  {journal} {J
  Chem Phys}\ }\textbf {\bibinfo {volume} {129}},\ \bibinfo {pages} {124105}
  (\bibinfo {year} {2008})}\BibitemShut {NoStop}%
\end{thebibliography}%
\end{document}